\documentclass[multphys,vecphys]{svmult}

\usepackage{graphicx}    

\makeindex             

\usepackage{amsmath}
\usepackage{amssymb,amscd}
\usepackage[arrow,matrix,curve]{xy}
\usepackage{graphicx}

\def\={\ =\ }
\def\bb{{\rm b}}
\def\BB{{\rm B}}
\def\Br{{\rm Br}}

\newcommand{\bbr}{\mathbb{R}}
\newcommand{\bbR}{\mathbb{R}}
\newcommand{\bbc}{\mathbb{C}}

\newcommand{\bbS}{\mathbb{S}}

\DeclareMathOperator{\K}{K}

\DeclareMathOperator{\KK}{KK}
\DeclareMathOperator{\Hom}{Hom}
\DeclareMathOperator{\RH}{H}
\DeclareMathOperator{\E2}{E}
\newcommand{\HL}{{\rm HL}}
\DeclareMathOperator{\HC}{HC}
\DeclareMathOperator{\Fred}{Fred}
\DeclareMathOperator{\Hd}{HH}
\DeclareMathOperator{\HP}{HP}

\DeclareMathOperator{\End}{End}
\DeclareMathOperator{\Aut}{Aut}
\DeclareMathOperator{\Ext}{Ext}
\DeclareMathOperator{\tor}{tor}

\DeclareMathOperator{\ch}{ch}
\DeclareMathOperator{\Ind}{index}

\newcommand{\ball}{\mathbb{B}}

\newcommand{\xr}{\xrightarrow}

\def\alg{{\mathcal A}}
\def\balg{{\mathcal B}}
\def\calg{{\mathcal C}}
\def\dalg{{\mathcal D}}
\def\ealg{{\mathcal E}}

\def\hil{{\mathcal H}}
\def\bun{{\mathcal E}}

\def\cK{{\mathcal K}}

\def\ecat{{\sf E}}
\def\lfcat{{\sf LF}}

\def\op{{\rm o}}
\def\Id{{\rm id}}
\def\pt{{\rm pt}}
\def\Cl{{\rm Cliff}}
\def\ch{{\rm ch}}

\def\Todd{{\rm Todd}}

\newcommand{\bbz}{{\mathbb Z}}
\newcommand{\bbq}{{\mathbb Q}}
\newcommand{\bbt}{{\mathbb T}}
\newcommand{\bbT}{{\mathbb T}}

\newcommand{\calS}{{\mathcal S}}
\newcommand{\calL}{{\mathcal L}}

\def\Poin{{\mathcal P}}
\def\Pic{{\rm Pic}}
\def\Dirac{{D\!\!\!\!/\,}} 
\newcommand{\dtorus}{{\widehat{\bbt}{}}}

\newcommand{\mbf}[1]{{\boldsymbol {#1} }}

\newcommand{\complex}{{\mathbb C}} 
\newcommand{\zed}{{\mathbb Z}} 
\newcommand{\mat}{{\mathbb M}} 
\newcommand{\PP}{{\mathbb P}}

\newcommand{\id}{{1\!\!1}} 
\def\alg{{\cal A}}

\newcommand{\Tr}[1]{\:{\rm Tr}\,#1}

\def\beq{\begin{equation}}
\def\eeq{\end{equation}}
\def\bea{\begin{eqnarray}}
\def\eea{\end{eqnarray}}
\def\bd{\begin{displaymath}}
\def\ed{\end{displaymath}}

\def\dd{{\rm d}}

\def\ii{{\,{\rm i}\,}}

\begin{document}

\title*{D-BRANES AND BIVARIANT K-THEORY}

\titlerunning{D-branes and bivariant K-theory}
\author{Richard J. Szabo}
\institute{Department of Mathematics and Maxwell Institute for
  Mathematical Sciences, Heriot-Watt University, Colin Maclaurin
  Building, Riccarton, Edinburgh EH14 4AS, U.K.
\texttt{R.J.Szabo@ma.hw.ac.uk}}
%
%
\maketitle

\begin{minipage}{11cm}
\small \baselineskip=11pt

We review various aspects of the topological classification of D-brane
charges in K-theory, focusing on techniques from geometric K-homology
and Kasparov's KK-theory. The latter formulation enables an elaborate
description of D-brane charge on large classes of noncommutative
spaces, and a refined characterization of open string T-duality in
terms of correspondences and KK-equivalence. The examples of D-branes
on noncommutative Riemann surfaces and in constant $H$-flux
backgrounds are treated in detail. Mathematical constructions include
noncommutative generalizations of Poincar\'e duality and
K-orientation, characteristic classes, and the Riemann-Roch theorem.

\hfill

Based on invited lectures given at the workshop ``Noncommutative
  Geometry and Physics 2008 -- K-Theory and D-Brane --'', February
18--22 2008, Shonan Village Center, Kanagawa, Japan. To be
  published in the volume {\it Noncommutative Geometry and Physics
    III} by World Scientific.

\end{minipage}

\hfill

\hfill

\centerline{\small\sf HWM--08--10 , EMPG--08--16 ; September 2008}

\setcounter{equation}{0}
\section{Introduction}
\label{sec:1}

The subject of this paper concerns the intriguing relationship between
D-branes and K-theory. As is by now well-known, D-brane charges in
string theory are classified by the K-theory of the spacetime
$X$~\cite{MM1}--\cite{RS1}, or equivalently (in the absence of $H$-flux) by the $\K$-theory of the $C^*$-algebra $C_0(X)$ of continuous functions on $X$ vanishing at infinity. D-branes are sources for Ramond-Ramond
fields, which are differential forms on spacetime and are
correspondingly classified by a smooth refinement of K-theory called
the differential K-theory of $X$~\cite{MW1}--\cite{SV1}. This
topological classification has been used to explain a variety of
effects in string theory that ordinary homology or cohomology alone
cannot explain, such as the existence of stable non-BPS branes with
torsion charges, the self-duality and quantization of Ramond-Ramond
fields, and the appearence of certain subtle worldsheet anomalies and
Ramond-Ramond field phase factors in the string theory path
integral. It has also been used to predict many novel phenomena such
as the instability of D-branes wrapping non-contractible cycles, and
obstructions to the simultaneous measurement of electric and magnetic
Ramond-Ramond fluxes.

The classification of D-branes can be posed as the following
problem. Given a closed string background $X$ (a Riemannian spin
manifold with possibly other form fields), find all possible states of
D-branes in $X$. At the worldsheet level, these states are described
as consistent boundary conditions in an underlying boundary
superconformal field theory. However, many of these states have no
geometrical description. It has therefore proven useful in a variety
of contexts to regard D-branes as objects in a suitable category. The
classic example of this is in conjunction with topological string
theory and Kontsevich's homological mirror symmetry conjecture, in
which B-model D-branes live in a bounded derived category of coherent
sheaves, while A-model D-branes are objects in a certain Fukaya
category~\cite{Douglas1}. A more recent example has been used to
clarify the relationship between boundary conformal field theory and
K-theory, and consists in regarding open string boundary conditions in
the category of a two-dimensional open/closed topological field
theory~\cite{MS1}. In the following we will argue that when one
combines the worldsheet description with the target space
classification in terms of Fredholm modules, one is led to regard
D-branes as objects in a certain category of separable
$C^*$-algebras~\cite{BMRS1}. This is the category underlying Kasparov's
\emph{bivariant K-theory} (or KK-theory), and it is related to the
open string algebras which arise in string field 
theory~\cite{BMRS2,BMRS3}.

The advantages of using the bivariant extension of K-theory are
abundant and will be described thoroughly in what follows. It unifies
the K-theory and K-homology descriptions of D-branes. It possesses an
intersection product which provides the correct framework for
formulating notions of duality between generic separable
$C^*$-algebras, such as Poincar\'e duality. This can be used to
explain the equivalence of the K-theory and K-homology descriptions of
D-brane charge. It also leads to a new characterization of open string
T-duality as a certain categorical KK-equivalence, which refines and
generalizes the more commonly used characterizations in terms of
Morita equivalence~\cite{Schwarz1}--\cite{SW1}. 
The formalism is also well equipped to deal with
examples of ``non-geometric'' backgrounds which have appeared recently
in the context of flux compactifications~\cite{STW1}. In certain
instances, the noncommutative spacetimes can be viewed~\cite{GSN1} as
globally defined open string versions of Hull's
\emph{T-folds}~\cite{Hull1}, which are backgrounds that fail to be
globally defined Riemannian manifolds but admit a local description in
which open patches are glued together using closed string T-duality
transformations. KK-theory also provides us with a noncommutative
version of K-orientation, which generalizes the Freed-Witten anomaly
cancellation condition~\cite{FW1} and enables us to select the
consistent sets of D-branes from our category. Finally, bivariant
K-theory yields a noncommutative version of the D-brane charge
vector~\cite{MM1}.

In formulating the notions of D-brane charge and Ramond-Ramond fields on arbitrary $C^*$-algebras, one is faced with the problem of developing Poincar\'e duality and constructing characteristic classes in these general settings. From the mathematical perspective of
noncommutative geometry alone, the formalism thus enables us to develop more
tools for dealing with noncommutative spaces in the purely algebraic
framework of separable $C^*$-algebras. These include noncommutative
versions of Poincar\'e duality and orientation, topological invariants
of noncommutative spaces such as the Todd genus, and a noncommutative
version of the Grothendieck-Riemann-Roch theorem which is intimately
tied to the formulation of D-brane charge.

\section{D-branes and K-homology}
\label{sec:2}

We will begin by explaining the topological classification of D-branes 
using techniques of geometric K-homology~\cite{BD1,Jakob1},
following~refs.~\cite{RS1,RSV1}. In this setting, brane
charges are expressed in terms of the Chern character in
$\K$-homology formulated topologically by the Baum-Douglas
construction. Using the Fredholm module description available in analytic K-homology, this will lead to a description of brane charges later on more complicated spaces, in particular on noncommutative spacetime manifolds. Earlier work in this context can be found in~refs.~\cite{AST1,Szabo1}. 

\subsection{D-branes and K-cycles}
\label{sec:2.1}

Throughout this paper we will work in the context of Type~II
superstring theory. Let $X$ be a compact spin$^c$-manifold, with no
background $H$-flux (we will explain in detail later on what we mean
precisely by this condition). A D-brane in $X$ may then be defined to
be a Baum-Douglas K-cycle $(W,E,f)$~\cite{BD1}, where
$f:W\hookrightarrow X$ is a closed spin$^c$ submanifold called the
worldvolume of the brane, and $E\rightarrow W$ is a complex vector
bundle with connection called the Chan-Paton gauge bundle. The crucial
feature about the Baum-Douglas construction is that $E$ defines a
\emph{stable} element of the K-theory group $\K^0(W)$. The set of all
K-cycles forms an additive category under disjoint union. 

The quotient of the set of all K-cycles by Baum-Douglas ``gauge
equivalence'' is isomorphic to the K-homology of $X$, defined as the
collection of stable homotopy classes of Fredholm modules over the
commutative $C^*$-algebra $\alg=C(X)$ of continuous functions on
$X$. The isomorphism sends a K-cycle $(W,E,f)$ to the unbounded
Fredholm module $(\hil,\rho,\Dirac_E^{(W)})$, where
$\hil=L^2(W,S\otimes E)$ is the separable Hilbert space of square
integrable $E$-valued spinors on $W$, $\rho(\phi)=m_{\phi\circ f}$ is
the $*$-representation of $\phi\in\alg$ on $\hil$ by pointwise
multiplication with the function $\phi\circ f$, and $\Dirac_E^{(W)}$
is the $E$-twisted Dirac operator associated to the spin$^c$ structure
on $W$. The K-homology class $[W,E,f]$ of a D-brane depends only on
the K-theory class $[E]\in\K^0(W)$ of its Chan-Paton
bundle~\cite{RS1}. Actually, to make this map surjective one has to
work with more general K-cycles wherein $W$ is not necessarily a
submanifold of spacetime. We will return to this point later on. 

It follows that D-branes naturally provide K-homology classes on $X$,
dual to K-theory classes $f_!(E)\in\K^d(X)$, where $f_!$ is the
K-theoretic Gysin map and $d=\dim(X)-\dim(W)$ is the codimension of
the brane worldvolume in spacetime. The natural $\zed_2$-grading on
K-homology $\K_\bullet(X)$ is by parity of dimension $\dim(W)=p+1$,
and the K-cycle $(W,E,f)$ then corresponds to a
D$p$-brane. Following~ref.~\cite{RS1}, we will now describe the
Baum-Douglas gauge equivalence relations explicitly, together with
their natural physical interpretations. 

\paragraph{Bordism} 

Two K-cycles $(W_1,E_1,f_1)$ and $(W_2,E_2,f_2)$ are said to be
bordant if there exists a K-cycle with boundary $(M,E,f)$ such that 
$$
\big(\partial M\,,\,E|_{\partial M}\,,\,f|_{\partial M}\big)~\cong~
\big(W_1\amalg(-W_2)\,,\,E_1\amalg E_2\,,\,f_1\amalg f_2\big) \ ,
$$
where $-W_2$ denotes the manifold $W_2$ with the opposite spin$^c$
structure on its tangent bundle $TW_2$. If $X$ is locally compact,
this relation generates a boundary condition which guarantees that
D-branes have finite energy. In particular, it ensures that any
K-cycle $(W,E,f)$ is equivalent to the closed string vacuum
$(\emptyset,\emptyset,\emptyset)$ (with no D-branes) at ``infinity''
in $X$.

\paragraph{Direct sum}

If $E_i$, $i=1,2$ are complex vector bundles over $W$, then we
identify the K-cycles 
$$
(W,E_1\oplus E_2,f)\sim(W,E_1,f)\amalg(W,E_2,f) \ .
$$
This relation reflects gauge symmetry enhancement for coincident
branes. The bundle $E=\bigoplus_i\,E_i$ is the Chan-Paton bundle
associated to a bound state of D-branes with Chan-Paton bundles
$E_i\rightarrow W$, bound by open string excitations given by classes
of bundle morphisms $[\phi_{ij}]\in\Hom(E_i,E_j)$. Other open string
degrees of freedom correspond to classes in $\Ext^p(E_i,E_j)$,
$p\geq1$. 

\paragraph{Vector bundle modification}

Let $(W,E,f)$ be a K-cycle and let $F\rightarrow W$ be a real spin$^c$ 
vector bundle of rank $2n$, with associated bundles of Clifford
modules $S_0(F),S_1(F)\rightarrow W$ and their pullbacks
$S_\pm(F)\rightarrow F$ of rank $2^{n-1}$. Clifford multiplication
induces a bundle map $\sigma:S_+(F)\rightarrow S_-(F)$ which is an
isomorphism outside of the zero section. If $\id^\bbR$ denotes the
trivial real line bundle over $W$, then upon choosing a Hermitean
metric on the fibres of $F$ we can define the unit sphere bundle 
\beq
\widehat{W}~:=~ \bbS\big(F\oplus\id^\bbR\big)\ \cong \
\ball_+(F)\cup_{\bbS(F)}\ball_-(F)
\label{unitspherebun}\eeq
with bundle projection 
$$\pi\,:\,\widehat{W}~\longrightarrow~ W \ , $$
where $\ball_\pm(F)$ are two copies of the unit ball bundle $\ball(F)$
of $F$ whose boundary is the unit sphere bundle $\bbS(F)$. We can glue
$\calS_\pm(F)=S_\pm(F)\big|_{\ball_\pm}$ together by Clifford
multiplication to define the bundle 
$$
H(F)\=\calS_+(F)\cup_\sigma\calS_-(F) \ .
$$
The restriction $H(F)\big|_{\pi^{-1}(w)}$ is the Bott generator of the
$2n$-dimensional sphere $\pi^{-1}(w)=\bbS^{2n}$ for each $w\in W$. 
We impose the equivalence relation
$$
(W,E,f)~\sim~\big(\,\widehat{W}\,,\,H(F)\otimes\pi^*(E)\,,\,f\circ
\pi\big) \ ,
$$
where the right-hand side is called the vector bundle modification of
$(W,E,f)$ by $F$. 

This relation can be understood as the K-homology version of the
well-known dielectric effect in string theory~\cite{Myers1}. To
understand this point, consider the simple K-cycle
$(W,E,f)=(\pt,\bbc,\iota)$, where $\iota$ is the inclusion of the
point $\pt$ in $X$. Let $F=\bbr^{2n}$, $n\geq1$. Then, with the
definitions above, one has $\widehat{W}\cong\bbS^{2n}$ with
$\pi:\bbS^{2n}\rightarrow\pt$ the collapsing map
$\varepsilon$. Moreover, $H(F)=H(F)\big|_{\bbS^{2n}}$ is the Bott
generator of the K-theory group $\K^0(\bbS^{2n})$. By vector bundle
modification, one has an equality of classes of K-cycles given by 
$$
[\pt,\bbc,\iota]\=\big[\bbS^{2n}\,,\,H(F)\otimes\bbc\,,\,\iota\circ
\pi\big] \=\big[\bbS^{2n}\,,\,H(F)\,,\,\varepsilon\big] \ .
$$
This equality represents the polarization or ``blowing up'' of a
D0-brane (on the left) into a collection of spherical D$(2n)$-branes
(on the right), together with ``monopole'' gauge fields corresponding
to connections on the vector bundles $H(F)\to\bbS^{2n}$. It is
essentially the statement of Bott periodicity.

\subsection{Tachyon condensation and the Sen-Witten construction}
\label{sec:2.2}

The Sen-Witten construction~\cite{Witten1,Sen1} is the classic model
establishing that D-brane charge is classified by K-theory. It relies
on the physics of tachyon condensation and the realization of stable
D-branes as decay products in unstable systems of spacetime filling
branes and antibranes. In K-homology, this construction utilizes the
fact that not all K-cycles are associated with submanifolds of
spacetime, and correspond generically to non-representable D-branes
arising as conformal boundary conditions with no direct geometric
realization. 

For definiteness, let $X$ be a locally compact spin manifold of
dimension $\dim(X)=10$. Let $W\subset X$ be a spin$^c$ submanifold of
dimension $p+1$. Then the normal bundle $\nu_W\rightarrow W$
to $W$ in $X$ is a real spin$^c$ vector bundle of rank $9-p$. A
D-brane $[M,E,\phi]\in\K_\bullet(X)$ is said to \emph{wrap} $W$ if
$\dim(M)=p+1$ and $\phi(M)\subset W$. The group of charges of Type~IIB  
D$p$-branes ($p$ odd) wrapping $W$ may then be computed as the
compactly supported K-theory group
\begin{eqnarray}
\K^0(\nu_W)&:=&\K^0\big(\ball(\nu_W)\,,\,\bbS(\nu_W)\big)
\nonumber \\[4pt]&\cong&\K_{10}(\nu_W) \nonumber \\[4pt]
&\cong&\K_{p+1}(W) \ , 
\label{IIBiso}\end{eqnarray}
where the first isomorphism follows from Poincar\'e duality and the
second from the K-homology Thom isomorphism. Upon identifying the
total space of $\nu_W$ with a tubular neighbourhood of $W$ in $X$ with 
respect to a chosen Riemannian metric on $X$, the group
$\K_{10}(\nu_W)$ classifies 9-branes in $X$. The isomorphism
(\ref{IIBiso}) then asserts that this group coincides with the group
of D$p$-branes $[M,E,\phi]$ wrapping $W$. The same calculation carries
through for Type~IIA D$p$-branes with $p$ even, starting from the
pertinent K-theory group $\K^{-1}(\nu_W)$. The original
example~\cite{Witten1,Sen1} concerns the charge group of D$p$-branes
in Type~IIB string theory on flat space $X=\bbR^{10}$ given by
$$
\K_{p+1}(\bbR^{p+1})~\cong~\K_0(\bbR^{10})~:=~
\widetilde{\K}_0(\bbS^{10}) \ ,
$$
where we have used Bott periodicity.

To make this relationship more explicit, we can adapt the
\emph{Atiyah-Bott-Shapiro (ABS) construction}~\cite{ABS1} to the
setting of geometric K-homology. Given a K-cycle $(W,E,f)$ in $X$, the 
vector bundle modification relation for $F=\nu_W$ reads
$$
\big[\,\widehat{W}\,,\,H(\nu_W)\otimes\pi^*(E)\,,\,f\circ\pi\big]\=
[W,E,f]
$$
with $\widehat{W}$ diffeomorphic to $X$. Generally, the nowhere
vanishing section given by $s:W\rightarrow F\oplus\id^\bbR$,
$x\mapsto0_x\oplus1$ induces a Gysin homomorphism on K-theory
$s_!:\K^\bullet(W)\rightarrow\K^\bullet\big(\,\widehat{W}\,\big)$ with 
$$
s_!(E)\=\big[\pi^*(E)\otimes H(F)\big]
$$
by the K-theory Thom isomorphism. Let
$W'\cong\ball(\nu_W)\setminus\bbS(\nu_W)$ be a tubular neighbourhood
of $f(W)$ with closure $\overline{W}$, retraction $\rho:W'\rightarrow
W$, and twisted spinor bundles
$\calS_E^\pm:=\calS^\pm(\nu_W)\otimes\rho^*(E)\rightarrow\overline{W}$. After 
a possible K-theoretic stabilization, we can extend the spinor bundles
over the complement $X\setminus W'$ to bundles $\calS_E^\pm\rightarrow
X$ with K-theory class~\cite{Witten1,OS1,ABS1}
$$
\big[\calS_E^+\big]-\big[\calS_E^-\big]\=s_!(E) \ ,
$$
which vanishes over $X\setminus W'$ by Clifford multiplication. Putting
everything together finally gives
$$
\big[X\,,\,\calS_E^+\,,\,\Id_X\big]-\big[X\,,\,\calS_E^-\,,\,\Id_X\big]
\=\pm\,[W,E,f] \ ,
$$
where the sign depends on whether or not the spin$^c$ structures on
$\widehat{W}$ and $X$ coincide. This equation is simply the statement
of tachyon condensation on the unstable spacetime-filling
brane-antibrane system (on the left) to a stable D-brane wrapping $W$
(on the right).

\subsection{Holonomy on D-branes}
\label{sec:2.3}

In order to cancel certain worldvolume anomalies, it is necessary to
introduce Ramond-Ramond flux couplings in the path integral for
Type~II string theory~\cite{DMW1}. The formalism of geometric
K-homology nicely achieves this via a spin$^c$ cobordism invariant as
follows. Introduce ``background'' D-branes $(W,E,f)$ as
\emph{K-chains}
$\big(\,\widetilde{W}\,,\,\widetilde{E}\,,\,\widetilde{f}\,\big)$ with
boundary
$$
\partial\big(\,\widetilde{W}\,,\,\widetilde{E}\,,\,\widetilde{f}\,
\big)~:=~\big(\partial\widetilde{W}\,,\,\widetilde{E}\,
\big|_{\partial\widetilde{W}}\,,\,
\widetilde{f}\,\big|_{\partial\widetilde{W}}\big)\=(W,E,f) \ .
$$
By bordism, such branes have trivial K-homology class and so carry no
charge.

The reduced eta-invariant of a K-chain is defined by
$$
\Xi\big(\,\widetilde{W}\,,\,\widetilde{E}\,,\,\widetilde{f}\,\big)\=
\mbox{$\frac12$}\,\big(\dim\hil_E^{(W)}+\eta(\Dirac_E^{(W)})\big)~\in~
\bbR/\bbz \ ,
$$
where $\hil_E^{(W)}$ is the space of harmonic $E$-valued spinors on
$W$, and $\eta\big(\Dirac_E^{(W)}\big)$ is the (regulated) spectral
asymmetry of the Dirac operator $\Dirac_E^{(W)}$. This invariant is defined up to compact perturbation of the Dirac operator and hence is $\bbR/\bbz$-valued. The map $\Xi$ from
K-chains to the group $\bbR/\bbz$ respects disjoint union, direct sum
and vector bundle modification, but \emph{not} spin$^c$ cobordism. To
rectify this problem, we introduce the holonomy over the given D-brane
background with flat Ramond-Ramond flux
$\xi=[E_0]-[E_1]\in\K^{-1}(X,\bbR/\bbz)\cong\Hom\big(\K_{\rm
  odd}(X)\,,\,\bbR/\bbz\big)$ by
$$
\Omega\big(\,\widetilde{W}\,,\,\widetilde{\xi}\,,\,\widetilde{f}\,
\big)\=\exp\Big[2\pi\ii\Big(\Xi\big(\,\widetilde{W}\,,\,
\widetilde{f^*E_0}\,,\,\widetilde{f}\,\big)-\Xi
\big(\,\widetilde{W}\,,\,
\widetilde{f^*E_1}\,,\,\widetilde{f}\,\big)\Big)\Big] \ .
$$
This quantity is the desired spin$^c$ cobordism invariant.

\subsection{Brane stability}
\label{sec:2.4}

We will now illustrate some of the predictive power of the K-homology
classification through two novel sets of examples of brane
stability which contradict what ordinary homology theory alone
would predict. The first set consists of trivial K-homology classes
$[W,\id^\bbc,f]=0$ in $\K_\bullet(X)$, even though the worldvolume
homology cycle $[W]\neq0$ in $\RH_\bullet(X,\bbz)$. The obstructions
to extending the homology class $[W]$ to a K-homology class are
measured by the Atiyah-Hirzebruch-Whitehead spectral sequence
$$
\E2_{p,q}^2\=\RH_p\big(X\,,\,\K_q(\pt)\big)\quad \Longrightarrow
\quad\K_{p+q}(X) \ .
$$
With respect to a cellular decomposition of the spacetime manifold
$X$, for each $p$ the corresponding filtration groups classify
D-branes wrapping $W$ on the $p$-skeleton of $X$ with no lower brane
charges.

The $r$-th term in the spectral sequence is determined as the
homology of certain differentials $\dd^r$. Cycles for which
$[W]\notin\ker(\dd^r)$ for all $r$ correspond to Freed-Witten
anomalous D-branes~\cite{MMS1}. On the other hand, if
$[W]\in{\rm im}(\dd^r)$ for some $r$, then the K-homology ``lift'' of
the cycle $[W]$ vanishes and the D-brane is unstable. Cycles contained
in the image of $\dd^r$ correspond to D-brane instantons~\cite{MMS1}
whose charge is not conserved in time along the trajectories of the
worldvolume renormalization group flow. The extension problem for the
spectral sequence at each term identifies the lower brane charges
carried by stable D-branes.

The second set is opposite in character to the first in that now
$[W,E,f]\neq0$ in $\K_\bullet(X)$ even though $[W]=0$ in
$\RH_\bullet(X,\bbz)$. This occurs by the process of flux
stabilization~\cite{MMS1}--\cite{BRS1} on spacetimes which are the
total spaces of topologically non-trivial fibre bundles
$X\xrightarrow{F}B$. Worldvolume ``flux'' in this instance
corresponds to the characteristic class of the fibration, which
provides a conserved charge preventing the D-brane from decaying to
the vacuum. Although $W$ 
is contractible in $X$, its class may be non-trivial as an element of
$\RH_\bullet(B,\bbz)$. The obstructions to lifting homology cycles
from the base space $B$ are measured by the Leray-Serre spectral sequence 
$$
\E2_{p,q}^2\=\RH_p\big(B\,,\,\K_q(F)\big) \quad \Longrightarrow \quad
\K_{p+q}(X,F) \ .
$$

Let us examine the original example of this phenomenon, that of
D-branes in the group manifold of $SU(2)$~\cite{BDS1}, in this
language. Spacetime in this case is the total space of the Hopf
fibration $\bbS^3\xrightarrow{\bbS^1}\bbS^2$, and the spectral
sequence computes the K-homology as $\K_i(X,\bbS^1)\cong
\RH_2(\bbS^2,\bbz)=\bbz$. The stable branes are spherical D2-branes,
and the stabilizing flux is provided by the first Chern class of the
monopole line bundle over $\bbS^2$. This example readily generalizes
to the other Hopf fibrations~\cite{RS1}, and the K-homology framework
nicely extends the examples of~refs.~\cite{MMS1,BRS1} to spaces with
less symmetry.

\subsection{D-brane charges}
\label{sec:2.5}

We will now describe the cohomological formula for the charge of a
D-brane~\cite{MM1}. The mathematical structure of this formula can be
motivated by the following simple observation, which we will
generalize later on to certain classes of noncommutative
spacetimes. The natural bilinear pairing in cohomology is given by
\beq
(x,y)_{\RH}\=\big\langle x\,\smile\,y\,,\,[X]\big\rangle
\label{cohpairing}\eeq
for cohomology classes $x,y\in\RH^\bullet(X,\zed)$ in complimentary
degrees. Upon choosing de~Rham representatives $\alpha,\beta$ for
$x,y$, this formula corresponds to integration of the product of
differential forms $\int_X\,\alpha\wedge\beta$. Nondegeneracy of this
pairing is the statement of Poincar\'e duality in cohomology. On the
other hand, the natural bilinear pairing in K-theory is provided on
complex vector bundles $E,F\to X$ by the index of the twisted Dirac
operator
\beq
(E,F)_{\K}\=\Ind(\Dirac_{E\otimes F}) \ .
\label{indexpairing}\eeq
The $\bbz_2$-graded Chern character ring isomorphism
\beq
\ch\,:\,\K^\bullet(X)\otimes\bbq~\xrightarrow{\approx}~
\RH^\bullet(X,\bbq)
\label{chiso}\eeq
is not compatible with these two pairings. However, by the
Atiyah-Singer index theorem
\beq
\Ind(\Dirac_{E\otimes F}) \= \big\langle \Todd (X) \smile
\ch(E\otimes F)\,,\, [X] \big\rangle
\label{ASindexthm}\eeq
we get an isometry with the $\bbz_2$-graded modified Chern character group isomorphism
$$
\ch~\longrightarrow~\sqrt{\Todd(X)}\,\smile\,\ch \ ,
$$
twisted by the square root of the invertible Todd class
$\Todd(X)\in\RH^{\rm even}(X,\bbq)$ of the tangent bundle $TX$. 

This almost trivial observation motivates the definition of the
\emph{Ramond-Ramond charge} of a D-brane $(W,E,f)$ as~\cite{MM1}
\beq
Q(W,E,f)\=\ch\big(f_!(E)\big)\,\smile\,\sqrt{\Todd(X)}~\in~
\RH^\bullet(X,\bbq) \ .
\label{RRcharge}\eeq
In topological string theory, this rational charge vector coincides
with the zero mode part of the associated boundary state in the
Ramond-Ramond sector. In the D-brane field theory,
${Q}(W,E,f)=f_*\big({D}_{\rm WZ}(W,E,f)\big)$ is the cohomological
Gysin image of the \emph{Wess-Zumino class} (for vanishing $B$-field) 
\beq
{D}_{\rm WZ}(W,E,f)\=\ch(E)\smile\sqrt{\Todd(W)/\,\Todd(\nu_W)}~\in~
\RH^\bullet(W,\bbq) \ .
\label{WZclass}\eeq
This formula interprets the Ramond-Ramond charge as the anomaly inflow
on the D-brane worldvolume $W$. The equivalence of these two formulas
follows from the Grothendieck-Riemann-Roch formula
\beq
\ch\big(f_!(E)\big) \smile {\Todd}(X) \= f_*\big(
\ch(E)\smile {\Todd}(W)\big)
\label{GRRthm}\eeq
together with naturality of the Todd characteristic
class. Compatibility with the equivalence relations of geometric
K-homology follows easily by direct calculation. In particular,
invariance under vector bundle modification is a simple computation
showing that the charge of the polarized D-brane
$\big(\,\widehat{W}\,,\,s_!(E)\,,\,f\circ\pi\big)$ equals $Q(W,E,f)$.

\section{D-branes and KK-theory}
\label{sec:3}

By merging the worldsheet and target space descriptions of D-branes,
we will now motivate a categorical framework for the classification of
D-branes using Kasparov's KK-theory groups. This will set the stage
for a noncommutative description of D-branes in a certain category of
separable $C^*$-algebras. We will then explain various important
features of the bivariant version of K-theory, and use them for
certain physical and mathematical constructions. The material of this
section is based on~refs.~\cite{BMRS1}--\cite{BMRS3}.

\subsection{Algebraic characterization of D-branes}
\label{sec:3.1}

The worldsheet description of a D-brane with worldvolume $W\subset X$
is provided by open strings, which may be defined to be relative maps
$(\Sigma,\partial\Sigma)\rightarrow(X,W)$ from an oriented Riemann
surface $\Sigma$ with boundary $\partial\Sigma$. In the boundary
conformal field theory on $\Sigma=\bbr\times[0,1]$, solutions of the
Euler-Lagrange equations require the imposition of suitable boundary
conditions, which we will label by $a,b,\dots$. These boundary
conditions are not arbitrary and compatibility with superconformal
invariance severely constrains the possible worldvolumes $W$. For
example, in the absence of background $H$-flux, $W$ must be spin$^c$
in order to ensure the cancellation of global worldsheet
anomalies~\cite{FW1}. The problem which now arises is that while this
is more or less understood at the classical level, there is no
generally accepted definition of what is meant by a \emph{quantum}
D-brane. Equivalently, it is not known in general how to define
consistent boundary conditions after quantization of the boundary
conformal field theory. To formulate our conjectural description of
this, we will take a look at the generic structure of open string
field theory. 

The basic observation is that the concatenation of open string vertex
operators defines algebras and bimodules. An $a$-$a$ open string, one
with the same boundary condition $a$ at both of its ends, defines a
noncommutative algebra $\dalg_a$ of open string fields. The opposite
algebra $\dalg_a^\op$, with the same underlying vector space as
$\dalg_a$ but with the product reversed, is obtained by reversing the
orientation of the open string. On the other hand, an $a$-$b$ open
string, with generically distinct boundary conditions $a,b$ at its two
ends, defines a $\dalg_a$-$\dalg_b$ bimodule $\ealg_{ab}$, with
the rule that open string ends can join only if their boundary labels
are the same. The dual bimodule $\ealg_{ab}^\vee=\ealg_{ba}$ is
obtained by reversing orientation, and $\ealg_{aa}=\dalg_a$ is defined
to be the trivial $\dalg_a$-bimodule on which $\dalg_a$ acts by (left
and right) multiplication.

We would now like to use these ingredients to define a ``category of
D-branes'' whose objects are the boundary conditions, and whose
morphisms $a\to b$ are precisely the bimodules $\ealg_{ab}$. This
requires an associative $\bbc$-bilinear composition law
$$
\ealg_{ab}\times\ealg_{bc}~\longrightarrow~\ealg_{ac} \ .
$$
The problem, however, in the way that we have set things up, is that
the operator product expansion of the open string fields is not always
well-defined. Elements of the open string bimodule $\ealg_{ab}$ are
vertex operators
$$
V_{ab}\,:\,[0,1]~\longrightarrow~\End(\hil_{ab})
$$ 
acting on a separable Hilbert space $\hil_{ab}$. The structure of the
vertex operator algebra is encoded in the singular operator product
expansion 
$$
V_{ab}(t)\cdot V_{bc}(t'\,)\=\sum_{j=1}^N\,\frac1{(t-t'\,)^{h_j}}~
W_{abc;j}(t,t'\,) \ , \quad t>t' \ ,
$$
where $W_{abc;j}:[0,1]\times[0,1]\to\End(\hil_{ac})$ and $h_j\geq0$ are
called conformal dimensions. When $h_j>0$, the leading singularities
of the operator product expansion do not give an associative algebra
in the usual sense.

\subsection{Seiberg-Witten limit}
\label{sec:3.2}

Seiberg and Witten~\cite{SW1} found a resolution to this problem in
the case where spacetime is an $n$-dimensional torus $X=\bbt^n$ with a
constant $B$-field. They introduced a scaling limit wherein both the
$B$-field and the string tension $T$ are scaled to infinity in such a
way that their ratio $B/T$ remains finite, while the closed string
metric $g$ on $\bbt^n$ is scaled to zero. In this limit the Hilbert
space $\hil_a$ of the point particle at an open string endpoint is a
module for a noncommutative torus algebra $\dalg_a$, which forms the
complete set of observables for boundary conditions of maximal
support. The product $\dalg_a\otimes\dalg_b$ acts irreducibly on the
$\dalg_a$-$\dalg_b$ bimodule $\ealg_{ab}=\hil_a\otimes\hil_b^\vee$.

In this case, the composition law
$$
V_{ac}(t'\,)\=\lim_{t\to t'}\,V_{ab}(t)\cdot V_{bc}(t'\,)
$$
is well-defined since the conformal dimensions scale to zero in the
limit as $h_j\sim g/T\to0$. It extends by associativity of the
operator product expansion in the limit to a map
$$
\ealg_{ab}\otimes_{\dalg_b}\ealg_{bc}~\longrightarrow~\ealg_{ac} \ .
$$
Furthermore, there are natural identifications of algebras
$\dalg_a\cong\ealg_{ab}\otimes_{\dalg_b}\ealg_{ba}$ and
$\dalg_b\cong\ealg_{ba}\otimes_{\dalg_a}\ealg_{ab}$. These results all
mean that $\ealg_{ab}$ is a Morita equivalence bimodule, reflecting a
\emph{T-duality} between the noncommutative tori $\dalg_a$ and
$\dalg_b$.

\subsection{KK-theory}
\label{sec:3.3}

The construction of Section~\ref{sec:3.2} above motivates a
conjectural framework in which to move both away from the dynamical
regime dictated by the Seiberg-Witten limit and into the quantum
realm. We will suppose that the appropriate modification consists in
replacing $\ealg_{ab}$ by Kasparov bimodules $(\ealg_{ab},F_{ab})$,
which generalize Fredholm modules. They coincide with the ``trivial''
bimodule $(\ealg_{ab},0)$ when $\ealg_{ab}$ is a Morita equivalence
bimodule. We will not enter into a precise definition of these
bimodules, which is somewhat technically involved
(see~ref.~\cite{BMRS1}, for example). As we move our way deeper into
our treatment we will become better acquainted with the structures
inherent in Kasparov's theory.

Stable homotopy classes of Kasparov bimodules define the
$\bbz_2$-graded KK-theory group
$\KK_\bullet(\dalg_a,\dalg_b)$. Classes in this group can be thought
of as ``generalized'' morphisms $\dalg_a\rightarrow\dalg_b$, in a way 
that we will make more precise as we go along. In particular, if
$\phi:\alg\rightarrow\balg$ is a homomorphism of separable
$C^*$-algebras, then it determines a canonical class
$[\phi]\in\KK_\bullet(\alg,\balg)$ represented by the ``Morita-type''
bimodule $(\balg,\phi,0)$. The group
$\KK_\bullet(\alg,\bbc)=\K^\bullet(\alg)$ is the K-homology of the
algebra $\alg$, since in this case Kasparov bimodules are the same
things as Fredholm modules over $\alg$. On the other hand, the group
$\KK_\bullet(\bbc,\balg)=\K_\bullet(\balg)$ is the K-theory of
$\balg$.

One of the most powerful aspects of Kasparov's theory is the existence
of a bilinear, associative \emph{composition} or \emph{intersection
  product} 
$$
\otimes_{\balg}\,:\,\KK_i(\alg,\balg)\times\KK_j(\balg,\calg)~
\longrightarrow~\KK_{i+j}(\alg,\calg) \ .
$$
We will not attempt a general definition of the intersection product,
which is notoriously difficult to define. Later on we will see how it
is defined on specific classes of $C^*$-algebras. The product is
compatible with the composition of morphisms, in that if
$\phi:\alg\rightarrow\balg$ and $\psi:\balg\rightarrow\calg$
are homomorphisms of separable $C^*$-algebras then
$$
[\phi]\otimes_\balg[\psi]\=[\psi\circ\phi] \ .
$$

The intersection product makes $\KK_0(\alg,\alg)$ into a unital ring
with unit $1_\alg=[\Id_\alg]$, the class of the identity morphism on
$\alg$. It can be used to define Kasparov's bilinear, associative
\emph{exterior product}
\begin{eqnarray*}
\otimes \,:\,
\KK_i(\alg_1, \balg_1) \times \KK_j(\alg_2, \balg_2) &
\longrightarrow & \KK_{i+j}(\alg_1\otimes \alg_2, \balg_1\otimes 
\balg_2) \ , \\[4pt]
x_1\otimes x_2 &=& (x_1 \otimes 1_{\alg_2} )\otimes_{\balg_1\otimes
  \alg_2} (1_{\balg_1}\otimes x_2) \ .
\end{eqnarray*}
This definition also uses \emph{dilation}. If
$x=[\phi]\in\KK_j(\alg,\balg)$ is the class of the morphism
$\phi:\alg\rightarrow\balg$, then
$x\otimes1_\calg=[\phi\otimes\Id_\calg]\in
\KK_j(\alg\otimes\calg,\balg\otimes\calg)$ is the class of the
morphism $\phi\otimes\Id_\calg:\alg\otimes\calg\rightarrow
\balg\otimes\calg$.

The KK-theory groups have some nice properties, described
in~refs.~\cite{Higson1}--\cite{Meyer1}, which enable us to define our
\emph{D-brane categories}. There is an additive category whose objects
are separable $C^*$-algebras and whose morphisms $\alg\to\balg$ are
precisely the classes in $\KK_\bullet(\alg,\balg)$. This category is a
\emph{universal} category, in the sense that $\KK_\bullet(-,-)$ can be
characterized as the unique bifunctor on the category of separable
$C^*$-algebras and $*$-homomorphisms which is homotopy invariant,
compatible with stabilization of $C^*$-algebras, and respects split
exactness. The composition law in this category is provided by the
intersection product. The category is not abelian, but it is
\emph{triangulated}, like other categories of relevance in D-brane
physics. It further admits the structure of a ``weak'' monoidal
category, with multiplication given by the spatial tensor product on
objects, the external Kasparov product on morphisms, and with identity
the one-dimensional $C^*$-algebra $\bbc$. A diagrammatic calculus in
this tensor category is developed in~refs.~\cite{BMRS1,BMRS2}.

\subsection{KK-equivalence}
\label{sec:3.4}

As our first application of the bivariant version of K-theory, we
introduce the following notion which will be central to our treatment 
later on. Any given fixed element $\alpha\in\KK_d(\alg,\balg)$
determines homomorphisms on K-theory and K-homology given by taking intersection products
$$
\otimes_\alg\alpha\,:\,\K_j(\alg)~\longrightarrow~\K_{j+d}(\balg)
\qquad \mbox{and} \qquad
\alpha\otimes_\balg\,:\,\K^j(\balg)~\longrightarrow~\K^{j+d}(\alg) \ .
$$
If $\alpha$ is \emph{invertible}, i.e., there exists an element
$\beta\in\KK_{-d}(\balg,\alg)$ such that
$\alpha\otimes_\balg\beta=1_\alg$ and
$\beta\otimes_\alg\alpha=1_\balg$, then we write $\beta=:\alpha^{-1}$
and there are isomorphisms
$$
\K_j(\alg)~\cong~\K_{j+d}(\balg) \qquad \mbox{and}
\qquad\K^j(\balg)~\cong~\K^{j+d}(\alg) \ .
$$
In this case the algebras $\alg,\balg$ are said to be
\emph{KK-equivalent}. From a physical perspective, algebras
$\alg,\balg$ are KK-equivalent if they are isomorphic as objects in
the D-brane category described at the end of Section~\ref{sec:3.3}
above. Such D-branes have the same K-theory and K-homology.

For example, Morita equivalence implies KK-equivalence, since the
discussion of Section~\ref{sec:3.2} above shows that the element
$\alpha=\big[(\bun_{ab},0)\big]$ is invertible. However, the converse
is not necessarily true. On the class of $C^*$-algebras which are
KK-equivalent to commutative algebras, one has the universal
coefficient theorem for KK-theory given by the exact
sequence~\cite{RSch1} 
\begin{eqnarray}
0 &\longrightarrow&{\Ext_\bbz}\bigl(\K_{\bullet+1}(\alg)\,,\,
\K_\bullet(\balg)\bigr) ~\longrightarrow~ \KK_\bullet\bigl(\alg\,,\,
\balg\bigr)~\longrightarrow ~ \nonumber \\ &\longrightarrow&
{\Hom_\bbz}\bigl(\K_\bullet(\alg)\,,\,
\K_\bullet(\balg) \bigr)~\longrightarrow~ 0 \ .
\label{UCTKK}\end{eqnarray}
We will make extensive use of this exact sequence in the following.

\subsection{Poincar\'e duality}
\label{sec:3.5}

The noncommutative version of Poincar\'e duality was introduced by
Connes~\cite{Connes1} and further developed
in~refs.~\cite{KP1}--\cite{Tu1}. Our treatment is closest to that of
Emerson~\cite{Emerson1}. Let $\alg$ be a separable $C^*$-algebra, and
let $\alg^\op$ be its opposite algebra. The opposite algebra is
introduced in order to regard $\alg$-bimodules as
$(\alg\otimes\alg^\op)$-modules. We say that $\alg$ is a
\emph{Poincar\'e duality (PD) algebra} if there is a \emph{fundamental 
  class} $\Delta\in\KK_d(\alg\otimes\alg^\op,\bbc)=
\K^d(\alg\otimes\alg^\op)$ with inverse
$\Delta^\vee\in\KK_{-d}(\bbc,\alg\otimes\alg^\op)=
\K_{-d}(\alg\otimes\alg^\op)$ such that
\begin{eqnarray*}
\Delta^\vee\otimes_{\alg^\op}\Delta&=&1_\alg\in\KK_0(\alg,\alg) \ ,
\\[4pt]
\Delta^\vee\otimes_{\alg}\Delta&=&(-1)^d~
1_{\alg^\op}\in\KK_0(\alg^\op,\alg^\op)
\end{eqnarray*}
for some $d=0,1$. The subtle sign factor in this definition reflects the orientation of the Bott element $\Delta^\vee$.

This definition determines inverse isomorphisms
\begin{eqnarray*}
\K_i(\alg)&\xrightarrow{\otimes_\alg\Delta}&\K^{i+d}(\alg^\op)=
\K^{i+d}(\alg) \ ,  \\[4pt] \K^i(\alg)=\K^{i}(\alg^\op)&
\xrightarrow{\Delta^\vee\otimes_{\alg^\op}}&
\K_{i-d}(\alg) \ ,
\end{eqnarray*}
which is the usual requirement of Poincar\'e duality. More generally,
by replacing the opposite algebra $\alg^\op$ in this definition with
an arbitrary separable $C^*$-algebra $\balg$, we get the notion of
\emph{PD pairs} $(\alg,\balg)$. Although the class of PD algebras is
quite restrictive, PD pairs are rather abundant~\cite{BMRS2}.

As a simple example, consider the commutative algebra
$\alg=C_0(X)=\alg^\op$ of continuous functions vanishing at infinity
on a complete oriented manifold $X$. Let $\balg=C_0(T^*X)$ or
$\balg=C_0\big(X\,,\,\Cl(T^*X)\big)$, where $T^*X$ is the cotangent
bundle over $X$ and $\Cl(T^*X)$ is the Clifford algebra bundle of
$T^*X$. Then $(\alg,\balg)$ is a PD pair, with $\Delta$ given by the
Dirac operator on $\Cl(T^*X)$. If in addition $X$ is a spin$^c$
manifold, then $\alg$ is a PD algebra. In this case, the fundamental
class $\Delta$ is the Dirac operator $\Dirac$ on the diagonal of
$X\times X$, i.e., the image of the Dirac class
$[\Dirac]\in\K^\bullet(\alg)$ under the group homomorphism
$$
m^*\,:\,\K^\bullet(\alg)~\longrightarrow~\K^\bullet(\alg\otimes\alg)
$$ 
induced by the product homomorphism
$m:\alg\otimes\alg\rightarrow\alg$, while its inverse $\Delta^\vee$ is
the Bott element. Thus in this case the noncommutative version of
Poincar\'e duality agrees with the classical one. We will encounter
some purely noncommutative examples later on. See ref.~\cite{BMRS1} for further examples.

In general, the moduli space of fundamental classes of an algebra
$\alg$ is isomorphic to the group of invertible elements in the ring
$\KK_0(\alg,\alg)$~\cite{BMRS1}. When $\alg=C_0(X)$, this space is in
general larger than the space of spin$^c$ structures or K-orientations 
usually considered in the literature. This follows from the universal
coefficient theorem (\ref{UCTKK}), which shows that the moduli space
is an extension of the automorphism group ${\rm
  Aut}(\K^0(X))$. Similarly, if $\alg$ and $\balg$
are $C^*$-algebras that are KK-equivalent, then the space of all
KK-equivalences $\alpha$ is a torsor with associated group the
invertible elements of $\KK_0(\alg,\alg)$. 

\subsection{K-orientation and Gysin homomorphisms}
\label{sec:3.6}

We can treat generic K-oriented maps by generalizing a construction
due to Connes and Skandalis in the commutative case~\cite{CSk1}. Let
$f:\alg\rightarrow\balg$ be a $*$-homomorphism of separable
$C^*$-algebras in a suitable category. Then a \emph{K-orientation} for
$f$ is a functorial way of associating a class
$f!\in\KK_d(\balg,\alg)$ for some $d=0,1$. This element determines a
\emph{Gysin ``wrong way'' homomorphism} on K-theory through
$$
f_!=\otimes_\balg (f!)\,:\,\K_\bullet(\balg)~\longrightarrow~
\K_{\bullet+d}(\alg) \ .
$$

If the $C^*$-algebras $\alg$ and $\balg$ are both PD algebras with
fundamental classes
$\Delta_\alg\in\KK_{d_\alg}(\alg\otimes\alg^\op,\bbc)$ and
$\Delta_\balg\in\KK_{d_\balg}(\balg\otimes\balg^\op,\bbc)$,
respectively, then any morphism $f:\alg\rightarrow\balg$ is
K-oriented with K-orientation given by 
$$
f!\=(-1)^{d_\alg}~\Delta_\alg^\vee\otimes_{\alg^\op}[f^\op]
\otimes_{\balg^\op}\Delta_\balg
$$
and $d=d_\alg-d_\balg$. This construction uses the fact~\cite{BMRS1}
that the involution $\alg\to\alg^\op$, $f\mapsto
f^\op:\alg^\op\to\balg^\op$ on the stable homotopy category of
separable $C^*$-algebras and $*$-homomorphisms passes to the
D-brane category. Functoriality
$$
g!\otimes_\balg f!=(g\circ f)!
$$
for any other $*$-homomorphism of separable $C^*$-algebras
$g:\balg\to\calg$ follows by associativity of the Kasparov
intersection product. More general constructions of K-orientations
will be encountered later on. 

The following construction demonstrates that any D-brane $(W,E,f)$ in
$X$ determines a canonical KK-theory class
$f!\in\KK_d\big(C(W)\,,\,C(X)\big)$. Recall that in this instance the
normal bundle $\nu_W=f^*(TX)/TW$ is a spin$^c$ vector bundle. Let
$i^W!:=\big[(\ealg,F)\big]\in\KK_d\big(C(W)\,,\,C(X)\big)$ be the
invertible element associated to the ABS representative of the Thom
class of the zero section $i^W:W\hookrightarrow\nu_W$. Let
$[\Phi]\in\KK_0\big(C_0(\nu_W)\,,\,C_0(W'\,)\big)$ be the invertible
element induced by the isomorphism $\Phi$ identifying $W'$ with a
neighbourhood of $i^W(W)$ in $X$. Let
$j!\in\KK_0\big(C_0(W'\,)\,,\,C(X)\big)$ be the class induced by the
extension by zero of the open subset $j:W'\hookrightarrow X$. Then a
K-orientation for $f$ is given by 
$$
f!\=i^W!\otimes_{C_0(\nu_W)}[\Phi]\otimes_{C_0(W'\,)}j! \ .
$$
 In this way our notion of K-orientation coincides with the
 Freed-Witten anomaly cancellation condition~\cite{FW1}. This
 construction extends to arbitrary smooth proper maps
 $\phi:M\rightarrow X$, corresponding generally to non-representable
 D-branes, for which $TM\oplus\phi^*(TX)$ is a spin$^c$ vector bundle
 over $M$.

\section{Cyclic theory}
\label{sec:4}

The definition of D-brane charge given in Section~\ref{sec:2.5} relied
crucially on the connection between the topological $\K$-theory of a
spacetime $X$ and its cohomology through the rational isomorphism
provided by the Chern character~(\ref{chiso}). In the generic
noncommutative settings that we are interested in, we need a more
general cohomological framework in which to express the D-brane
charge. The appropriate receptacle for the Chern character in analytic
$\K$-theory is the cyclic cohomology of the given (noncommutative)
algebra $\alg$~\cite{Connes1}. As it is not commonly known material in
string theory, in this section we will present a fairly detailed
overview of the general aspects of cyclic homology and
cohomology. Then we will specialize to the specific bivariant cyclic
theory that we will need in subsequent sections. This general
formulation will provide a nice intrinsic definition of the D-brane
charge, suitable to our noncommutative situations.

\subsection{Cyclic homology}
\label{sec:4.1}

Let $\alg$ be a unital algebra over $\complex$. The {\it universal
  differential graded algebra $\Omega^\bullet(\alg)$} is the universal
algebra generated by $\alg$ and the symbols $\dd a$, $a\in\alg$
with the following properties:
\begin{enumerate}
\item $\dd:\alg\to\Omega^1(\alg)$ is linear;
\item $\dd$ obeys the Leibniz rule $\dd(a\,b)=\dd(a)~b+a~\dd(b)$;
\item $\dd(1)=0$; and
\item $\dd^2=0$.
\end{enumerate}
These conditions imply that $\dd$ is a linear derivation, and elements
of $\Omega^\bullet(\alg)$ are called {\it noncommutative differential
  forms} on $\alg$, or more precisely on the tensor algebra
$T\alg=\bigoplus_{n\geq0}\,\alg^{\otimes n}$ of $\alg$. We define
$\Omega^0(\alg)=\alg$. In degree $n>0$, the space of $n$-forms is the
linear span
$$
\Omega^n(\alg)={\rm Span}_\complex\big\{a_0~\dd a_1\cdots\dd a_n~
\big|~a_0,a_1,\dots,a_n\in\alg\big\} \ ,
$$
which under the isomorphism $a_0~\dd a_1\cdots\dd a_n\leftrightarrow
a_0\otimes a_1\otimes\cdots\otimes a_n$ may be presented explicitly as 
a vector space by
$$
\Omega^n(\alg)\cong\alg\otimes(\alg/\complex)^{\otimes n} \ .
$$
The graded vector space $\Omega^\bullet(\alg)$ then becomes a graded
algebra by using the Leibniz rule to define multiplication of forms by 
\begin{eqnarray}
&& (a_0~\dd a_1\cdots\dd a_n)\cdot(a_{n+1}~\dd a_{n+2}\cdots\dd a_p)
\label{formmultdef} \\ && \qquad \= (-1)^n\,a_0\,a_1~\dd a_2\cdots\dd
a_p+\sum_{i=1}^n\,(-1)^{n-i}\,a_0~
\dd a_1\cdots\dd(a_i\,a_{i+1})\cdots\dd a_p \ . \nonumber
\end{eqnarray}
Using this definition the operator $\dd$ may be
extended to a graded derivation on $\Omega^\bullet(\alg)$.

When the algebra $\alg$ is not unital, we apply the above construction
to the unitization  $\widetilde{\alg}=\alg\oplus\complex$ of
$\alg$, with multiplication given by
$$
(a,\lambda)\cdot(b,\mu)=(a\,b+\lambda\,b+\mu\,a,\lambda\,\mu) \ .
$$
Thus in degree $n>0$ we have 
\beq
\Omega^n(\alg)~:=~\Omega^n\big(\,\widetilde{\alg}\,\big)
\=\alg^{\otimes(n+1)}\oplus \alg^{\otimes n} \ . 
\label{Omeganonunital}\eeq
In degree~$0$ we define $\Omega^0({\alg}) = \alg$.

The cohomology of the differential $\dd$ on $\Omega^\bullet(\alg)$ is
trivial in positive degree and equal to $\complex$ in degree $0$. To
get interesting homology theory, we need to introduce two other
differentials. Let us first define the boundary map
$$
\bb\,:\,\Omega^n(\alg)~\longrightarrow~\Omega^{n-1}(\alg)
$$
by the formula
$$
\bb (\omega ~\dd a) = (-1)^{|\omega|} \,[\omega, a]
$$
where $|\omega|=n-1$ is the degree of the form
$\omega\in\Omega^{n-1}(\alg)$. This definition uses the structure
of a differential graded algebra on $\Omega^\bullet
(\alg)$. Using the explicit  formula (\ref{formmultdef}) for the
product of two forms and assuming that $\omega = a_0~\dd a_1 \cdots
\dd a_{n-1}$, this definition may be rewritten in the form
\begin{eqnarray}
\bb(a_0\otimes a_1\otimes\cdots\otimes
a_n)&=&\sum_{i=0}^{n-1}\,(-1)^i\,a_0\otimes\cdots
\otimes a_i\,a_{i+1}\otimes\cdots\otimes
a_n\nonumber\\ &&+\,(-1)^n\,a_n\,a_0\otimes
a_1\otimes\cdots\otimes a_{n-1} \ .
\label{bbdef}\end{eqnarray}

The Karoubi operator is the degree~$0$ operator  $\kappa: \Omega^n 
(\alg) \to \Omega^n( \alg)$ defined by
$$
\kappa (\omega~ \dd a) = (-1)^{|\omega|} ~\dd a~\omega
$$
where $\omega \in \Omega^{n-1}(\alg)$. Explicitly, this operator is
given by the formula on
$\kappa:\alg^{\otimes(n+1)}\to\alg^{\otimes(n+1)}$ through
$$
\kappa(a_0\otimes a_1\otimes\cdots\otimes a_n)=
(-1)^n\,a_n\otimes a_0\otimes\cdots\otimes a_{n-1}+(-1)^n\,
1\otimes a_n\,a_0\otimes\cdots\otimes a_{n-1} \ .
$$
On the image $\dd\Omega^\bullet (\alg)$ of the differential $\dd$,
this operator is precisely  the generator (with sign) of cyclic
permutations. With this in mind we introduce the remaining
differential  
$$
\BB=\sum_{i=0}^n\,\kappa^i~\dd\,:\,\Omega^n(\alg)~\longrightarrow~
\Omega^{n+1}(\alg) \ .
$$

It is easy to check that the two operators $\bb$ and $\BB$ anticommute
and are nilpotent,
$$
\bb^2\=\bb~\BB+\BB~\bb\=\BB^2\=0 \ .
$$
The two differentials $\BB$ and $\bb$ give $\Omega^\bullet (\alg)$ the
structure of a \emph{mixed} complex  $(\Omega^\bullet( \alg), \bb,
\BB)$, which can be organised into a double complex given by the
diagram
\beq
\xymatrix{
 & \vdots\ar[d]_\bb & \vdots\ar[d]_\bb\ar[ld]^S& {}\ar[ld]^S\\
 \cdots~& 
~\Omega^{n+1}(\alg)~ \ar[l]^\BB\ar[d]_\bb\ar[ld]^S
&~\Omega^n(\alg)~\ar[l]^\BB\ar[d]_\bb\ar[ld]^S&~\cdots\ar[l]^\BB
\ar[ld]^S 
\\  \cdots~& ~\Omega^{n}(\alg)~ \ar[l]^\BB\ar[d]_\bb\ar[ld]^S
&~\Omega^{n-1}(\alg)~\ar[l]^\BB\ar[d]_\bb\ar[ld]^S&\cdots~ \ar[l]^\BB
\\ & \vdots & \vdots & }
\label{bicomplex}\eeq
which in bidegree $(p,q)$ contains $\Omega^{p-q}(\alg)$. The columns
in this complex are repeated and we declare all spaces located at
$(p,q)$ with $p-q < 0$ or $p<0$ to be trivial. Thus this double
complex occupies one octant in the $(p,q)$-plane. There is a canonical 
isomorphism $S$ which by definition is the identity map sending
the space $\Omega^n(\alg)$ located at $(p+1, q+1)$ to itself located
at $(p,q)$. The column at $p=0$ is by definition annihilated by
$S$. This operator is Connes' periodicity operator. It follows from
its definition that $S$ is of degree~$-2$. 

The {\it total complex $({\rm Tot}\,\Omega^\bullet(\alg),\bb+\BB)$} of
the bicomplex $(\Omega^\bullet(\alg),\bb,\BB)$ is defined in degree
$n$ by the finite sum
$$
{\rm
  Tot}_n\,\Omega^\bullet(\alg)=\bigoplus_{p\geq0}\,\Omega^{n-2p}(\alg)
\ . 
$$
The {\it Hochschild homology} $\Hd_\bullet(\alg)$ of the algebra
$\alg$ is defined to be the homology of the complex
$(\Omega^\bullet(\alg),\bb)$, 
\beq
\Hd_\bullet(\alg)=\RH_\bullet\big(\Omega^\bullet(\alg)\,,\,\bb\big) \
. 
\label{HHdef}\eeq
The {\it cyclic homology $\HC_\bullet(\alg)$} of the algebra $\alg$
is defined to be the homology of the total complex $({\rm
  Tot}\,\Omega^\bullet(\alg),\bb+\BB)$, 
$$
\HC_\bullet(\alg)=\RH_\bullet\big({\rm
  Tot}\,\Omega^\bullet(\alg)\,,\,\bb+\BB\big) \ . 
$$
If we denote by $I: \Omega^\bullet (\alg) \to {\rm Tot}\,
\Omega^\bullet(\alg)$ the inclusion of the first column into the
double complex (\ref{bicomplex}), then by using the definition of the
Connes periodicity operator it is not difficult to deduce the
fundamental relation between Hochschild and cyclic homology given by
the long exact sequence
\begin{eqnarray}
\cdots~\longrightarrow~ \Hd_{n+2}(\alg)~
\stackrel{I}{\longrightarrow}~ \HC_{n+2}(\alg)~ &\stackrel{S}
{\longrightarrow}& ~\HC_{n}(\alg)~ \stackrel{\BB}{\longrightarrow}~
\label{homlongexact} \\ &\stackrel{\BB}{\longrightarrow}&~
\Hd_{n+1}(\alg)~ \longrightarrow~\cdots \ . \nonumber
\end{eqnarray}
The map $S$ in this sequence is induced by the periodicity operator
which gives rise to  a surjection $S:
\text{Tot}_{n+2}\,\Omega^\bullet(\alg) \to
\text{Tot}_{n}\,\Omega^\bullet(\alg)$.

Finally, we define the periodic cyclic homology. For this, we need to 
consider a complex that is a completion, in a certain sense, of the
total complex used in  the construction of cyclic homology. Thus we
put
$$
\widehat{\Omega}{}^\bullet(\alg) = \prod_{n\geq 0}\, \Omega^n(\alg) \ . 
$$
Elements of this space are inhomogenous forms $(\omega_0, \omega_1,
\dots , \omega_n, \dots)$, where $\omega_n\in  \Omega^n(\alg)$, with
possibly infinitely many non-zero components. We shall regard this
space as being $\bbz_2$-graded with the decomposition into even and
odd degree forms given by 
$$
\widehat{\Omega}^{\rm even}(\alg)\=\prod_{n\geq0}\,
{\Omega}^{2n}(\alg) \qquad \mbox{and} \qquad
\widehat{\Omega}^{\rm odd}(\alg)\=\prod_{n\geq0}\,
{\Omega}^{2n+1}(\alg) \ .
$$
A typical element of $\widehat{\Omega}^{\text{even}}(\alg)$ is a
sequence $(\omega_0, \omega_2, \dots ,
\omega_{2n}, \dots )$, and
similarly for $\widehat{\Omega}^{\text{odd}}(\alg)$. Then the {\it
  periodic cyclic homology} $\HP_\bullet(\alg)$ of the algebra $\alg$
is defined to be the homology of the $\zed_2$-graded complex
$$
\cdots~\xr{\bb+\BB}~\widehat{\Omega}^{\rm even}(\alg)~
\xr{\bb+\BB}~\widehat{\Omega}^{\rm odd}(\alg)~
\xr{\bb+\BB}~\widehat{\Omega}^{\rm even}(\alg)~
\xr{\bb+\BB}~\cdots \ .
$$

The Connes operator $S$ also provides a relation between the cyclic
and periodic cyclic homology in the following way. For every $n$,
there is a surjection
$$
T_{2n}\,:\, \widehat{\Omega}^{\text{even}}(\alg) ~\longrightarrow~
\text{Tot}_{2n}\,\Omega^\bullet (\alg)
$$
which sends a form $(\omega_0, \omega_2, \dots)$ to its 
truncation $(\omega_0, \omega_2, \dots, \omega_{2n})$. For various
values of $n$ these surjections are compatible with the periodicity
operator $S$ in the sense that there is a commutative diagram
\beq
\xymatrix{
&   \text{Tot}_{2n+2}\,\Omega^\bullet(\alg) \ar[dd]^S \\
\widehat{\Omega}^{\text{even}}(\alg)~
\ar[ur]^{T_{2n+2}} \ar[dr]_{T_{2n}} \\
&  \text{Tot}_{2n}\,\Omega^\bullet(\alg) \ . \\
}
\label{commdiagS}\eeq
An even periodic cycle is a sequence of the type described above which
is annihilated by the  operator $\bb + \BB$, i.e., applying
this operator creates the zero chain in
$\widehat{\Omega}^{\text{odd}} (\alg)$ as in the diagram
$$
\begin{CD}
\vdots\\
@V{\bb}VV \\
0 @<{\BB}<< \omega_{4} \\
@.      @V{\bb}VV \\
@.           0 @<{\BB}<< \omega_2 \\
@.  @. @V{\bb}VV \\
@.  @. 0 @<{\BB}<< \omega_0 \\
@.   @.  @. @V{0}VV \\
@. @.  @.  ~~ \ 0 \ .
\end{CD}
$$
The vertical map in degree~$0$ is the zero map. The truncation of this
cycle in, say, degree~$2$ creates an element of
$\text{Tot}_2\,\Omega^\bullet (\alg)$ which is a cycle in the
\emph{cyclic} complex
$$
\begin{CD}
   0 @<{0}<< \omega_2 \\
@.  @V{\bb}VV \\
@.  0 @<{\BB}<< \omega_0 \\
@.  @. @V{0}VV \\
@.  @.  ~~ \ 0 \ .
\end{CD}
$$
The zero map in the upper left corner appears due to the definition of
the differential in the cyclic complex. It kills the leftmost column
(where $p=0$), which in this case is the column where $\omega_2$ is
located.

Thus, for any $n$ the truncation map $T_{2n}$ sends an even periodic
cycle to a cyclic $2n$-cycle and so induces a map $T_{2n} :
\HP_{\rm even}(\alg) \rightarrow\HC_{2n}(\alg)$. From the diagram
(\ref{commdiagS}) it follows that these
maps are compatible with the periodicity operator $S$ and we obtain
a surjection
$$
\HP_{\rm even}(\alg) ~\longrightarrow~ \lim_{\stackrel{\scriptstyle
\longleftarrow}{\scriptstyle S}}\,\HC_{2n}(\alg) \ .
$$
There is a complementary map in odd degree whose construction is
identical to the one just described. This map is not an isomorphism in 
general. Its kernel is equal to 
$\displaystyle{\lim_{\longleftarrow}}{}^1\,\HC_{\bullet+2n+1}(\alg)$,
where $\displaystyle{\lim_{\longleftarrow}}{}^1$ is the first derived
functor of the inverse limit functor.

We will now consider a key example which 
illustrates the importance of these constructions. Let 
$\alg=C^\infty(X)$ be the algebra of smooth functions on a smooth
paracompact spacetime manifold $X$. Then the action of the boundary 
map (\ref{bbdef}) is trivial and the mixed complex
$(\Omega^\bullet(\alg),\bb,\BB)$ reduces to the complexified de~Rham
complex $(\Omega^\bullet (X),\dd)$, where $\dd$ is the usual de~Rham
exterior derivative on $X$. Equivalently, there is a natural
surjection
$\mu:(\Omega^\bullet(\alg),\bb,\BB)\to(\Omega^\bullet(X),0,\dd)$ of 
mixed complexes. The Connes-Pflaum version of the
Hochschild-Kostant-Rosenberg theorem asserts that the map $\mu$ is a
quasi-isomorphism, i.e., it induces equality of the Hochschild
homology (\ref{HHdef}) with the de~Rham complex. Explicitly, the map
$\mu :\Omega^n(\alg)\to\Omega^n(X)$ is implemented by sending a
noncommutative $n$-form to a differential $n$-form as 
$$
\mu(f^0~\dd f^1\cdots\dd f^n)=\mbox{$\frac1{n!}$}\,f^0~\dd
f^1\wedge\cdots\wedge\dd f^n
$$
for $f^i\in C^\infty(X)$. It follows that the Hochschild homology
of the algebra $C^\infty(X)$ gives the de~Rham complex,
$$
\Hd_n\bigl(C^\infty(X)\bigr)\cong\Omega^n(X) \ ,
$$
which implies that the periodic cyclic homology computes the periodic
de~Rham cohomology as
\beq
\HP_{\rm even}\bigl(C^\infty(X)\bigr)~\cong~\RH_{\rm dR}^{\rm
even}\bigl(X\bigr) \qquad \mbox{and} \qquad
\HP_{\rm odd}\bigl(C^\infty(X)\bigr)~\cong~
\RH_{\rm dR}^{\rm odd}\bigl(X\bigr) \ .
\label{HPHdR}\eeq
It is in this sense that we may regard cyclic homology
as a generalization of de~Rham cohomology to other (possibly
noncommutative) settings.

\subsection{Cyclic cohomology}
\label{sec:4.2}

As one would expect, by considering the duals of the chain spaces
introduced in Section~\ref{sec:4.1} above, one obtains the cohomology
theories corresponding to the three cyclic-type homology theories
defined there.  A {\it Hochschild $n$-cochain} on the algebra $\alg$
is a linear form on $\Omega^n (\alg)$, or equivalently an
$(n+1)$-multilinear functional $\varphi$ on $\alg$ which is
simplicially normalized in the sense that $\varphi(a_0, a_1, \dots,
a_n) = 0$ if $a_i = 1$ for any $i$ such that $1\leq i \leq n$. With 
the collection of all $n$-cochains denoted $C^n(\alg)={\rm
  Hom}_\bbc(\Omega^n(\alg),\complex)$, we form the {\it Hochschild
  cochain complex} $(C^\bullet(\alg),\bb)$ with coboundary map
$$
\bb\,:\, C^n(\alg)~\longrightarrow~C^{n+1}(\alg)
$$
given by the transpose of the differential $\bb$ as
\begin{eqnarray*}
\bb\varphi(a_0,a_1,\dots,a_{n+1})&=&\sum_{i=0}^n\,(-1)^i\,
\varphi(a_0,\dots,a_i\,a_{i+1},\dots,a_{n+1})\nonumber\\ &&+\,
(-1)^{n+1}\,\varphi(a_{n+1}\,a_0,a_1,\dots,a_n) \ .
\end{eqnarray*}
The cohomology of this complex is the {\it Hochschild
cohomology} 
$$
\Hd^\bullet(\alg)=\RH^\bullet\big(C^\bullet(\alg)\,,\,\bb\big) \ ,
$$
the dual theory to Hochschild homology defined in eq.~(\ref{HHdef}).

Similarly, the operator $\BB$ transposes to the cochain complex
$C^\bullet(\alg)$ and the {\it cyclic cohomology} $\HC^\bullet(\alg)$
is defined as the cohomology of the complex
$((\text{Tot}\,\Omega^\bullet(\alg))^\vee, \bb+\BB)$. The dual of the
periodic complex is the complex which in even degree is spanned by
\emph{finite} sequences $(\varphi_0, \varphi_2, \dots, \varphi_{2n})$
with $\varphi_i \in C^i(\alg)$, and similarly in odd degree. The {\it
  periodic cyclic cohomology} $\HP^\bullet(\alg)$ is the
cohomology of this complex. The long exact sequence
(\ref{homlongexact}) relating Hochschild and cyclic homology has an
obvious dual sequence that links Hochschild and cyclic cohomology. The
transpose of the periodicity operator provides an injection
$S:\HC^n(\alg)\to\HC^{n+2}(\alg)$ of cyclic cohomology groups and
therefore gives rise to two \emph{inductive} systems of abelian
groups, one running through even degrees and the other through odd
degrees. One has
$$
\HP^\bullet(\alg)  =
\lim_{\stackrel{\scriptstyle\longrightarrow}{\scriptstyle
S}}\,\HC^{\bullet+2n}(\alg) \ .
$$

This formal approach to cyclic cohomology, while very useful, hides
two important features  of the theory. Firstly, it seems to imply
that cyclic cohomology is secondary to cyclic homology. 
In fact, it turns out that many geometric and analytic situations
provide natural examples of cyclic \emph{cocycles}~\cite{Connes1}. 
Secondly, this approach does not explain why cyclic cohomology is
indeed cyclic. For this, we note that a Hochschild $0$-cocycle
$\tau\in{\rm Hom}(\alg,\complex)$ on the algebra $\alg$ is a trace,
i.e., $\tau(a_0\,a_1)=\tau(a_1\,a_0)$. This tracial property is
extended to higher orders via the following notion. Let
$\lambda:C^n(\alg)\to C^n(\alg)$ be the operator defined by
$$
\lambda\varphi(a_0,a_1,\dots,a_n)=(-1)^n\,\varphi(a_n,a_0,
\dots,a_{n-1}) \ .
$$
Then an $n$-cochain $\varphi\in C^n(\alg)$ is said to be {\it cyclic} 
if it is invariant under the action of the cyclic group,
$\lambda\varphi=\varphi$. The set of cyclic $n$-cochains is denoted
$C_\lambda^n(\alg)$. One can prove that the cohomology of this complex 
is isomorphic to the cohomology of the complex we have used above to
define cyclic cohomology, and so we can alternatively define the
{cyclic cohomology} $\HC^\bullet(\alg)$ of the algebra $\alg$ as the
cohomology of the cyclic cochain complex
$(C_\lambda^\bullet(\alg),\bb)$, 
$$
\HC^\bullet\big(\alg\big)=
\RH^\bullet\big(C^\bullet_\lambda(\alg)\,,\,\bb\big) \ .
$$

An important class of cyclic cocycles is obtained as follows. Consider 
the algebra $\alg = C^\infty(X)$ of smooth functions on a compact
oriented manifold $X$ of dimension $d$. Put 
\beq
\varphi^{~}_X\big(f^0\,,\,f^1\,,\, \dots\,,\, f^d\big) = \frac1{d!}\,
\int_X\, f^0~\dd f^1\wedge\dots\wedge \dd f^d
\label{varphiX}\eeq
for $f^i \in \alg$. Then $\varphi^{~}_X$ is a cyclic $d$-cocycle. More
generally, one can associate in this way a cyclic $(d-k)$-cocycle with
any closed $k$-current $C$ on $X$. In particular, the Chern-Simons
coupling $\big\langle C\,\smile\,Q(W,E,f)\,,\,[X]\big\rangle$ on a
D-brane $(W,E,f)$ is an inhomogeneous cyclic cocyle of definite parity 
for any closed cochain $C$ associated to a Ramond-Ramond field on
$X$.

\subsection{Local cyclic cohomology}
\label{sec:4.3}

Thus far we have not considered the possibility that the algebra
$\alg$ might be equipped with a topology. A major weakness of cyclic
cohomology compared to K-theory is that it depends very sensitively on
the domain of algebras. For instance, $\alg=C^\infty(X)$ is the
commutative, nuclear Fr\'{e}chet algebra of smooth functions on the
spacetime manifold $X$ equipped with its standard semi-norm
topology. More generally, we can allow $\alg $ to be a complete
multiplicatively convex algebra, i.e., $\alg$ is a topological algebra
whose topology is given by a family of submultiplicative
semi-norms. In such cases the definition of the
algebra $\Omega^\bullet(\alg)$ of noncommutative differential forms
will involve a choice of a suitably completed topological tensor
product $\overline{\otimes}$. The correct choice is forced by the
topology on $\alg$ and the corresponding continuity properties of the
multiplication map $m:\alg\otimes\alg\to\alg$. For nuclear Fr\'echet
algebras $\alg$, there is a unique topology which is compatible with
the tensor product structure on $\alg\otimes\alg$~\cite{BP1}. In our
later considerations we will often consider the situation in which
$\alg=\balg^\infty$ is a suitable smooth subalgebra of a separable
$C^*$-algebra $\balg$. Local cyclic cohomology is best suited to deal
with these and other classes of algebras, and it moreover has a useful
extension to a bivariant functor~\cite{Puschnigg1}. The bivariant
cyclic cohomology theories were introduced to provide a target for the
Chern character from KK-theory, which we describe in the next section.

The space of cochains in this theory is a certain deformation of the
space of maps $\Hom _\bbc(\,\widehat{\Omega}{}^\bullet (\alg),
\widehat{\Omega}{}^\bullet{(\balg)})$ with the $\bbz_2$-grading induced
from the spaces of inhomogeneous forms over the algebras $\alg$ and
$\balg$. Alternatively, we can define
$$
\Hom_\bbc\big(\,\widehat{\Omega}{}^\bullet(\alg)\,,\,
\widehat{\Omega}{}^\bullet(\balg)\,\big) = 
\lim_{\stackrel{\scriptstyle\longleftarrow}{\scriptstyle m}}~
\lim_{\stackrel{\scriptstyle\longrightarrow}{\scriptstyle n}}~
\Hom_\bbc \Big(\,\mbox{$\bigoplus\limits_{i\leq n}\,\Omega^i(\alg)
  \,,\, \bigoplus\limits_{j\leq m}\,\Omega^j(\balg)$}\Big) \ .
$$
This is a $\bbz_2$-graded vector space equipped with a differential
$\partial$ that acts on cochains $\varphi$ by the graded commutator
$$
\partial \varphi = [\varphi,  \bb + \BB] \ .
$$

The local version of this theory is defined by using a deformation of
the tensor algebra called the \emph{$X$-complex}, which is the
$\bbz_2$-graded completion of $\Omega^\bullet(\alg)$ given by 
$$
\xymatrix{
X^\bullet(T\alg)\,:\,\Omega^0(T\alg)=T\alg~
\ar@<0.5ex>[rr]^{\natural\circ\dd} &&
\ar@<0.5ex>[ll]^{\bb}~\Omega^1(T\alg)_\natural:=
\frac{\Omega^1(T\alg)}
{\big[\Omega^1(T\alg)\,,\,\Omega^1(T\alg)\big]} \ .
}
$$
Puschnigg's completion of $X^\bullet(T\alg)$~\cite{Puschnigg1},
$$
\xymatrix{
\widehat{X}{}^\bullet(T\alg)\,:\,\widehat{\Omega}^{\rm even}(\alg)~ 
\ar@<0.5ex>[rr] && \ar@<0.5ex>[ll]~
\widehat{\Omega}^{\rm odd}(\alg) \ ,
}
$$
then defines the $\bbz_2$-graded \emph{bivariant local cyclic
  cohomology} 
$$
\HL_\bullet(\alg,\balg)\=\RH_\bullet\big(\Hom_\bbc(\,
\widehat{X}{}^\bullet(T\alg),\widehat{X}{}^\bullet(T\balg)\,)\,,\,\partial\big)
\ . 
$$

The main virtues of Puschnigg's cyclic theory for our purposes is that
it is the one ``closest'' to Kasparov's KK-theory, in the sense that
it possesses the following properties. It is defined on large classes
of topological and bornological algebras, i.e., algebras together with
a chosen family of \emph{bounded} subsets closed under forming finite
unions and taking subsets, \emph{and} for separable $C^*$-algebras. It
defines a bifunctor $\HL_\bullet(-,-)$ which is homotopy invariant,
split exact and satisfies excision in each argument. It possesses a
bilinear, associative composition product 
$$
\otimes_\balg \,:\,
\HL_i(\alg,\balg)\times \HL_j(\balg,\calg) ~
\longrightarrow~ \HL_{i+j}(\alg,\calg) \ .
$$
It also carries a bilinear, associative exterior product
$$
\otimes \,:\,\HL_i(\alg_1,\balg_1)\times \HL_j(\alg_2,\balg_2) ~
\longrightarrow ~
\HL_{i+j}\big(\alg_1 \,\widehat{\otimes}\, \alg_2\,,\,
\balg_1 \,\widehat{\otimes}\, \balg_2\big) \ ,
$$
defined using the projective tensor product which maps onto the
minimal $C^*$-algebraic tensor product on the category of separable
$C^*$-algebras. In general, without any extra assumptions, this tensor
product differs from the usual spatial tensor product, but at least in
the examples we consider later on this problem can always be
fixed. Thus in what follows, we will not distinguish between the
algebraic tensor product $\otimes$ and its topological completion.

The local cyclic cohomology reduces to other cyclic theories under
suitable conditions, such as the periodic cyclic cohomology for
non-topological algebras and even Fr\'echet algebras, Meyer's analytic
theories for bornological algebras~\cite{Meyer2}, and Connes' entire
cyclic cohomology for Banach algebras. It thus possesses the same
algebraic properties as the usual bivariant cyclic cohomology
theories, and in this sense it unifies cyclic homology and
cohomology. A particularly useful property which we will make
extensive use of is the following. Let $\alg$ be a Banach algebra with
the metric approximation property, and let $\alg^\infty$ be a smooth
subalgebra of $\alg$. Then the inclusion
$\alg^\infty\hookrightarrow\alg$ induces an invertible element of
$\HL_0(\alg^\infty,\alg)$. Thus in this case the algebras
$\alg^\infty$ and $\alg$ are \emph{HL-equivalent}.

Let us consider again the illustrative example of the algebra of
functions $\alg=C(X)$ on a compact oriented manifold $X$ with
$\dim(X)=d$. In this case the inclusion of the smooth subalgebra
$C^\infty(X)\hookrightarrow C(X)$ gives an isomorphism~\cite{Meyer2}
$\HL\big(C(X)\big)\cong\HL\big(C^\infty(X)\big)\cong
\HP\big(C^\infty(X)\big)$ with the periodic cyclic cohomology. The
Puschnigg complex coincides with the periodic complexified de~Rham
complex $\big(\Omega^\bullet(X)\,,\,\dd\big)$. Using the isomorphism
(\ref{HPHdR}) we then arrive at the isomorphism of $\bbz_2$-graded
groups 
$$
\HL_\bullet\big(C(X)\big)~\cong~\RH_{\rm dR}^\bullet(X) \ .
$$
The cyclic $d$-cocycle (\ref{varphiX}) under this isomorphism induces
the orientation fundamental class
$\Xi=m^*[\varphi^{~}_X]\in\HL^d\big(C(X)\otimes C(X)\big)$
corresponding to the orientation cycle $[X]$ of the manifold $X$.

\section{D-brane charge on noncommutative spaces}
\label{sec:5}

In this section we will generalize the Minasian-Moore formula
(\ref{RRcharge}) for the Ramond-Ramond charge of a D-brane to large
classes of separable $C^*$-algebras representing generic
noncommutative spacetimes. This will require a few mathematical
constructions of independent interest on their own. In particular, we
will develop noncommutative versions of the characteristic classes
appearing in eq.~(\ref{RRcharge}) and show how they are related
through a generalization of the Grothendieck-Riemann-Roch theorem
(\ref{GRRthm}).

\subsection{Chern characters}
\label{sec:5.1}

We will begin by exhibiting the fundamental Chern character maps which 
link $\K$-theory and periodic cyclic homology, $\K$-homology and
periodic cyclic cohomology, and more generally $\KK$-theory and
bivariant cyclic cohomology. They provide explicit cyclic cocyles for
Fredholm modules, and establish crucial links between duality in
$\KK$-theory and in bivariant cyclic cohomology which will be the crux
of some of our later constructions. We begin with a description of the
Chern character in $\K$-theory. Let $\alg$ be a unital Fr\'echet
algebra over $\complex$. Acting on the $\K$-theory of the algebra
$\alg$, we construct the homomorphism of abelian groups
\beq
\ch_\sharp\,:\,\K_0(\alg)~\longrightarrow~\HP_{\rm even}(\alg)
\label{chK0HP0}\eeq
as follows. Let $[p]\in\K_0(\alg)$ be the Murray-von~Neumann
equivalence class of an idempotent
matrix $p\in\mat_r(\alg)=\alg\otimes\mat_r(\complex)$, i.e., a
projection $p=p^2$. Then the {\it Chern character} assigns to $[p]$ an
even class in the periodic cyclic homology of $\alg$ represented by
the even periodic cycle
$$
\ch_\sharp(p)=\Tr_r(p)+\sum_{n\geq1}\,\frac{(2n)!}{n!}\,\Tr_r\bigl(
(p-\mbox{$\frac12$})~\dd p^{2n}\bigr)
$$
valued in $\Omega^{\rm even}(\alg)$, where $\Tr_r:\mat_r(\alg)\to\alg$
is the ordinary $r\times r$ matrix trace. One readily checks that it
gives a cycle in the reduced $(\bb,\BB)$ bi-complex of cyclic homology
that we described in Section~\ref{sec:4.1}, i.e.,
$(\bb+\BB)\ch_\sharp(p)=0$. When $\alg=C^\infty(X)$ with $X$ a smooth
compact manifold, it coincides with the usual Chern-Weil character
$\Tr\exp(F/2\pi\ii)$ defined in terms of the curvature $F$ of the
canonical Grassmann connection of the corresponding complex vector
bundle $E\to X$. The Chern map (\ref{chK0HP0}) becomes an isomorphism
on tensoring with $\complex$.

For applications to the description of D-brane charges in cyclic
theory, it is more natural to use cyclic cohomology
classes corresponding to elements in $\K$-homology. Let
$(\hil,\rho,F)$ be an $(n+1)$-summable even Fredholm module
over the algebra $\alg$ with $n$ even. This means that
$[F,\rho(a)]\in\mathcal{L}^{n+1}$ for all $a\in\alg$, where
$\mathcal{L}^p=\mathcal{L}^p(\hil):=
\{T\in\mathcal{K}(\hil)~|~\Tr^{~}_\hil(T^p)<\infty\}$
is the $p$-class Shatten ideal of compact operators. Then the {\it
  character} of the Fredholm module is the cyclic $n$-cocycle $\tau^n$ 
given by
$$
\tau^n(a_0,a_1,\dots,a_n)=\Tr^{~}_\hil\bigl(\gamma\,\rho(a_0)\,
\bigl[F\,,\,
\rho(a_1)\bigr]\cdots\bigl[F\,,\,\rho(a_n)\bigr]\bigr) \ ,
$$
where $\gamma$ is the grading involution on $\hil$ defining its
$\zed_2$-grading into $\pm\,1$ eigenspaces of $\gamma$. One checks
closure $\bb\tau^n=0$ and cyclicity
$\lambda\tau^n=(-1)^n\,\tau^n$. Since 
$\mathcal{L}^{p_1}\subset\mathcal{L}^{p_2}$ for $p_1\leq p_2$, we can
replace $n$ by $n+2k$ with $k$ any integer in this definition, and
so only the (even) parity of $n$ is fixed. Thus for any $k\geq0$,
one gets a sequence of cyclic cocycles $\tau^{n+2k}$ with the same
parity. The cyclic cohomology classes of these cocycles are related by 
Connes' periodicity operator $S$ in $\HC^{n+2k+2}(\alg)$, and
therefore the sequence $(\tau^{n+2k})_{k\geq0}$ determines a
well-defined class $\ch^\sharp(\hil,\rho,F)$ called the {\it Chern
  character} of the even Fredholm module $(\hil,\rho,F)$ in the even
periodic cyclic cohomology $\HP^{\rm even}(\alg)$. Thus we get a map
$$
\ch^\sharp\,:\,\K^0(\alg)~\longrightarrow~\HP^{\rm even}(\alg)
$$
which becomes an isomorphism after tensoring over
$\complex$. See ref.~\cite{BMRS1} for an extension of this definition
to unbounded and infinite-dimensional Fredholm modules.

Our main object of interest is the Chern character in KK-theory. A
cohomological functor which complements the bivariant $\KK$-theory is
provided by the local bivariant cyclic cohomology
$\HL_\bullet(\alg,\balg)$ that we introduced in
Section~\ref{sec:4.3}. Since both $\KK_\bullet(\alg,\balg)$ and
$\HL_\bullet(\alg,\balg)$ are homotopy invariant, stable and satisfy
excision, the universal property of $\KK$-theory implies that there
is a natural bivariant $\bbz_2$-graded Chern character homomorphism
$$
\ch\,:\,\KK_\bullet(\alg,\balg)~\longrightarrow~\HL_\bullet(\alg,\balg)
$$
which enjoys the following properties:
\begin{enumerate}
\item $\ch$ is multiplicative, i.e., if $\alpha\in \KK_i(\alg,\balg)$
  and $\beta\in \KK_j(\balg,\calg)$ then
$$\ch(\alpha\otimes_\balg\beta) \= \ch (\alpha) \otimes_\balg
\ch(\beta) \ ; $$
\item $\ch$ is compatible with the exterior product; and
\item $\ch\big([\phi]_{\KK}\big)=[\phi]_{\HL}$ for any for any
  algebra homomorphism $\phi:\alg\rightarrow\balg$.
\end{enumerate}

The last property implies that the Chern character sends invertible
elements of $\KK$-theory to invertible elements of bivariant 
cyclic cohomology. In particular, every PD pair for KK-theory is also
a PD pair for HL-theory, but not conversely (due to
e.g. torsion). However, in the following it will be important to
consider distinct fundamental classes $\Xi\neq\ch(\Delta)$ in local
cyclic cohomology. If $\alg,\balg$ obey the universal coefficient
theorem (\ref{UCTKK}) for KK-theory, then there is an isomorphism
$$
\HL_\bullet(\alg,\balg)
~\cong~\Hom_\bbc\big(\K_\bullet(\alg)\otimes_{\bbz}\bbc \,,\,
\K_\bullet(\balg)\otimes_\bbz\bbc\big) \ .
$$
If the $\K$-theory $\K_\bullet(\alg)$ is finitely generated,
then this is also equal to
$$
\HL_\bullet(\alg,\balg)
~\cong~\KK_\bullet(\alg,\balg) \otimes_\bbz\bbc \ .
$$

\subsection{Todd classes}

Let $\alg$ be a PD algebra with fundamental K-homology class $\Delta\in\K^d(\alg\otimes\alg^\op)$, and fundamental cyclic cohomology class $\Xi\in\HL^d(\alg\otimes\alg^\op)$. Then we define the \emph{Todd class} of $\alg$ to be the element
$$
{\Todd(\alg)~:=~\Xi^\vee\otimes_{\alg^\op}\ch(\Delta)}~\in~\HL_0(\alg,\alg) \ .
$$
The Todd class is invertible with inverse given by
$$
\Todd(\alg)^{-1}\=(-1)^d~\ch\big(\Delta^\vee\big)\otimes_{\alg^\op}\Xi \ .
$$
More generally, one defines the Todd class $\Todd(\alg)$ for PD pairs of algebras $(\alg,\balg)$ by replacing $\alg^\op$ with $\balg$ above~\cite{BMRS1}.

The Todd class depends ``covariantly'' on the choices of fundamental classes in the respective moduli spaces~\cite{BMRS1}. For any other fundamental class $\Delta_1$ for $\K$-theory of $\alg$, one has
$ \Xi^\vee\otimes_{\alg^\op}{\rm ch}(\Delta_1)= {\rm
  ch}(\ell)\otimes_{\alg} {\Todd}(\alg)$
where $\ell = \Delta^\vee\otimes_{\alg^\op}\Delta_1 $ is an invertible element in
$\KK_0(\alg, \alg)$. Conversely, if $\ell$ is an invertible element in
$\KK_0(\alg, \alg)$, then $\ell \otimes_\alg \Delta$ is a fundamental
class for $\K$-theory of $\alg$ for any fundamental class $\Delta$. In particular, if $\alg,\balg$ are KK-equivalent $C^*$-algebras, with the KK-equivalence implemented by an invertible element $\alpha$ in $\KK_\bullet(\alg,\balg)$, then their Todd classes are related through
\beq
\Todd(\balg)\=\ch(\alpha)^{-1}\otimes_\alg\,\Todd(\alg)\otimes_\alg\,\ch(\alpha) \ .
\label{ToddKKrel}\eeq

The following example provides the motivation behind this definition. Let $\alg= C(X)$ where $X$ is a compact complex manifold. Then $\alg$ is a PD algebra, with KK-theory fundamental class $\Delta$ provided by the Dolbeault operator $\partial$ on $X\times X$, and HL-theory fundamental class $\Xi$ provided by the orientation cycle $[X]$. By the universal coefficient theorem (\ref{UCTKK}), one has an isomorphism $\HL_0(\alg,\alg)\cong\End\big(\RH^\bullet(X,\bbq)\big)$. Then $\Todd(\alg)=\,\smile\,\Todd(X)$ is cup product with the usual Todd characteristic class $\Todd(X)\in\RH^\bullet(X,\bbq)$ of the tangent bundle of $X$.

\subsection{Grothendieck-Riemann-Roch theorem}
\label{sec:5.3}

Let $f:\alg\rightarrow\balg$ be a K-oriented morphism of separable $C^*$-algebras. The Grothendieck-Riemann-Roch formula compares the class $\ch(f!)$ with the HL-theory orientation class $f*$ in $\HL_d(\balg,\alg)$. If $\alg$, $\balg$ are PD algebras, then one has $d=d_\alg-d_\balg$ and
\beq
{\ch(f!)\=(-1)^{d_\balg}~\Todd(\balg)\otimes_\balg(f*)\otimes_\alg
\Todd(\alg)^{-1}} \ .
\label{NCGRR}\eeq
This formula is proven by expanding out both sides using the various definitions, along with associativity of the Kasparov intersection product~\cite{BMRS1}. It leads to the commutative diagram
$$
\xymatrix{
\K_\bullet(\balg)~\ar[r]^{\!\!f_!}
\ar[d]_{\ch\otimes_\balg\Todd(\balg)} & 
~\K_{\bullet+d}(\alg)\ar[d]^{\ch\otimes_\alg\Todd(\alg)} \\
\HL_\bullet(\balg)~\ar[r]_{\!\!\!\!\!f_*} & ~
\HL_{\bullet+d}(\alg)
}
$$
generalizing eq.~(\ref{GRRthm}).

As an example of the applicability of this formula, suppose that $\alg$ is unital with even degree fundamental class. Then there is a canonical K-oriented morphism $\lambda:\bbc\rightarrow\alg$, $z\mapsto z\cdot1$ which induces a homomorphism on K-theory $\lambda_!:\K_0(\alg)\rightarrow\bbz$ with
\beq
\lambda_!(\xi) \= \lambda_*\bigl(\ch(\xi) \otimes_\alg
\Todd(\alg)  \bigr)
\label{lambdaK}\eeq
for $\xi\in\K_0(\alg)$. When $\alg=C(X)$, with $X$ a compact spin$^c$ manifold, then $\lambda_!(\xi)=\Ind(\Dirac_\xi)$ for $\xi\in\K^0(X)$ and eq.~(\ref{lambdaK}) is just the Atiyah-Singer index theorem (\ref{ASindexthm}). Generally, when $\xi=\alg$ is the trivial rank one module over $\alg$, then $\lambda_!(\xi)$ defines a characteristic numerical invariant of $\alg$, which we may call the \emph{Todd genus} of the algebra $\alg$.

\subsection{Isometric pairing formulas}

Suppose that $\alg$ is a PD algebra with \emph{symmetric} fundamental classes $\Delta$ and $\Xi$, i.e., $\sigma(\Delta)^\op=\Delta$ in $\K^d(\alg\otimes\alg^\op)$, where $\sigma:\alg\otimes\alg^\op\rightarrow\alg^\op\otimes\alg$ is the flip involution $x\otimes y^\op\mapsto y^\op\otimes x$, and similarly for $\Xi$ in $\HL^d(\alg\otimes\alg^\op)$. In this case we can define a symmetric bilinear pairing on the K-theory of $\alg$ by
\beq
{(\alpha,\beta)_{\K}\=\big(\alpha\otimes\beta^\op\big)\otimes_{\alg\otimes\alg^\op}\Delta}~\in~\KK_0(\bbc,\bbc)=\bbz
\label{NCindexpairing}\eeq
for $\alpha,\beta\in\K_\bullet(\alg)$. It coincides with the index pairing (\ref{indexpairing}) when $\alg=C(X)$, for $X$ a spin$^c$ manifold with fundamental class given by the Dirac operator $\Delta=\Dirac\otimes\Dirac$, as then
$$
(\alpha,\beta)_{\K} \=  \Dirac_\alpha \otimes_{C(X)} \beta \=
\Ind (\Dirac_{\alpha\otimes\beta})
$$
by definition of the intersection product on KK-theory. Similarly, one has a symmetric bilinear pairing on local cyclic homology given by
\beq
{(x,y)_{\HL}\=\big(x\otimes y^\op\big)\otimes_{\alg\otimes\alg^\op}\Xi}~\in~\HL_0(\bbc,\bbc)=\bbc \ ,
\label{HLpairing}\eeq
generalizing the pairing (\ref{cohpairing}).

If $\alg$ satisfies the universal coefficient theorem (\ref{UCTKK}), then one has an isomorphism $\HL_\bullet(\alg,\alg)\cong\End\big(\HL_\bullet(\alg)\big)$. If $\HL_\bullet(\alg)$ is a finite-dimensional vector space, $n:=\dim_{\bbc}\big(\HL_\bullet(\alg)\big)<\infty$, then we may use the universal coefficient theorem to identify the Todd class $\Todd(\alg)$ with an invertible matrix in $GL(n,\bbc)$. In this case the square root $\sqrt{\Todd(\alg)}$ may be defined using the usual Jordan normal form of linear algebra, and then reinterpreted as a class in $\HL_\bullet(\alg,\alg)$ again by using the universal coefficient theorem. This square root is not unique, but we assume that it is possible to fix a canonical choice. Under these circumstances, we can define the \emph{modified Chern character}
\beq
\ch\otimes_\alg\sqrt{\Todd(\alg)}\,:\,\K_\bullet(\alg)~
\longrightarrow~\HL_\bullet(\alg) \ ,
\label{modch}\eeq
which is an isometry of the inner products (\ref{NCindexpairing}) and (\ref{HLpairing})~\cite{BMRS1,BMRS2}.

Suppose now that $\alg$, $\dalg$ represent noncommutative D-branes with $\alg$ as above, with a given K-oriented morphism $f:\alg\rightarrow\dalg$ and Chan-Paton bundle $\xi\in\K_\bullet(\dalg)$. In this case there is a noncommutative version of Minasian-Moore formula (\ref{RRcharge}) given by
\beq
{Q(\dalg,\xi,f)\=\ch\big(f_!(\xi)\big)\otimes_\alg\sqrt{\Todd(\alg)}}~\in~\HL_\bullet(\alg) \ .
\label{NCRRcharge}\eeq
More generally, consider a D-brane in the noncommutative spacetime $\alg$ described by a Fredholm module over $\alg$ representing a K-homology class $\mu\in\K^\bullet(\alg)$. It has a ``dual'' charge given by
$$
{Q(\mu)\=\sqrt{\Todd(\alg)}{\,}^{-1}\otimes_\alg\ch(\mu)}~\in~\HL^\bullet(\alg) \ .
$$
This vector satisfies the isometry rule~\cite{BMRS1}
$$
\Xi^\vee\otimes_{\alg\otimes\alg^\op}\big(Q(\mu)\otimes Q(\nu)^\op\big)\=\Delta^\vee\otimes_{\alg\otimes\alg^\op}\big(\mu\otimes\nu^\op\big) \ ,
$$
and reproduces the noncommutative Minasian-Moore formula (\ref{NCRRcharge}) in the case when $\mu=f_!(\xi)\otimes_\alg\Delta$ is dual to the Chan-Paton bundle $\xi$.

\section{Noncommutative D2-branes} 

In this section we will apply our general formalism to the example of D-branes on noncommutative Riemann surfaces, as defined in refs.~\cite{CHMM1,Mathai1}. Consider a collection of D2-branes wrapping a compact, oriented Riemann surface $\Sigma_g$ of genus $g\geq1$ with a constant $B$-field. This example generalizes the classic example of D-branes on the noncommutative torus $\bbt_\theta^2$, obtained for $g=1$.

The fundamental group of $\Sigma_g$ admits the presentation
$$\Gamma_g \= \Bigl\{\mbox{$U_j, V_j\,,\, j=1, \ldots, g~
\Big|~\prod\limits_{j=1}^g\,[U_j, V_j] = 1$}\Bigr\} \ . $$
Its group cohomology is $\RH^2\big(\Gamma_g\,,\,U(1)\big)\cong\bbR/\bbz$, and so for each $\theta\in[0,1)$ there is a unique $U(1)$-valued two-cocyle $\sigma_\theta$ on $\Gamma_g$, representing the holonomy of the $B$-field on $\Sigma_g$. The reduced twisted group $C^*$-algebra $\alg_\theta:=C_r^*(\Gamma_g,\sigma_\theta)$ is isomorphic to the algebra generated by unitary elements $U_j,V_j$, $j=1,\dots,g$ obeying the single relation
$$
\prod_{j=1}^g \,[U_j, V_j] \= \exp (2\pi \ii\theta) \ .
$$
When $\theta$ is irrational, the degree~$0$ K-theory is $\K_0(\alg_\theta)\cong\K^0(\Sigma_g)=\bbz^2$, with generators $e_0=[1]$ and $e_1$ satisfying $\Tr(e_1)=\theta$, where $\Tr:C_r^*(\Gamma_g,\sigma_\theta)\rightarrow\bbc$ is the evaluation at the identity element $1_{\Gamma_g}$ of $\Gamma_g$. The degree~$1$ K-theory is given by $\K_1(\alg_\theta)\cong\K^1(\Sigma_g)=\bbz^{2g}$, with basis $U_j,V_j$. There is a smooth subalgebra $\alg_\theta^\infty\hookrightarrow\alg_\theta$ such that $\HL_\bullet(\alg_\theta)\cong\HL_\bullet(\alg_\theta^\infty) \cong\HP_\bullet(\alg_\theta^\infty)$~\cite{BMRS1,BMRS2}. 

The algebra $\alg_\theta$ is a PD algebra, with Bott class given by
$$
\Delta^\vee\=e^{\phantom{\op}}_0\otimes e^\op_1 - e^{\phantom{\op}}_1
\otimes e^\op_0 + \sum_{j=1}^g\,\left(U_j^{\phantom{\op}}
\otimes V^\op_j - V_j^{\phantom{\op}}\otimes U^\op_j\right) \ .
$$
Let $\mu_\theta :\K^\bullet(\Sigma_g) \xrightarrow{\approx}
\K_{\bullet}\big(C^*_r(\Gamma_g, \sigma_\theta)\big)$ be the twisted Kasparov isomorphism, and let $\nu_\theta$ be its analog in periodic cyclic homology. The commutative diagram of isomorphisms
$$
\xymatrix{
\K^\bullet(\Sigma_g) ~ \ar[r]^{\mu_\theta}\ar[d]_{\ch} & ~
\K_{\bullet}(\alg_\theta) \ar[d]^{\ch_{\Gamma_g}} \\
\RH^\bullet(\Sigma_g,\bbz)~ \ar[r]_{\nu_\theta} & ~ \HL_{\bullet}
(\alg_\theta)  }
$$
then serves to show that the Todd class is given by $\Todd(\alg_\theta)=\nu_\theta\big(\Todd(\Sigma_g)\big)$. This construction thus leads to the charge vector for a wrapped noncommutative D2-brane $(\dalg,\xi,f)$, with K-oriented morphism $f:\alg_\theta\rightarrow\dalg$ and Chan-Paton bundle $\xi\in\K_\bullet(\dalg)$, defined by
$$
{Q_\theta(\dalg,\xi,f) \= \nu_\theta\Big(
\ch\big(\mu_\theta^{-1}\circ
f_!(\xi)\big) \smile \sqrt{{\Todd}(\Sigma_g)}~\Big)}\in\HL_\bullet(\alg_\theta) \ .
$$
This formula incorporates the contribution from the constant $B$-field in the usual way~\cite{SW1,Taylor1}.

\section{D-branes and $\mbf H$-flux}
\label{sec:7}

In this section we will consider in some detail the example of D-branes in a compact, even-dimensional oriented manifold $X$ with constant background Neveu-Schwarz $H$-flux. In this case, it is well-known~\cite{Witten1,BM1} that one should replace spacetime $X$ by a noncommutative $C^*$-algebra $CT(X,H)$, the stable continuous trace $C^*$-algebra with spectrum $X$ and Dixmier-Douady invariant $H$~\cite{Rosenberg1}. This algebra has the property that it is locally Morita equivalent to spacetime, but not in general globally equivalent to it.

\subsection{Projective bundles and twisted K-theory}
\label{sec:7.1}

We will start by describing twisted K-theory, the appropriate receptacle for the classification of D-brane charge in $H$-flux backgrounds, in the spirit of Atiyah and Segal~\cite{AS1} (glossing over many topological details, as before). Let $\hil$ be a fixed, separable Hilbert space of dimension $\geq1$. We will denote the associated projective space of $\hil$ by $\PP=\PP(\hil)$. It is compact if and only if $\hil$ is finite-dimensional. Let $PU=PU(\hil)=U(\hil)/U(1)$ be the projective unitary group of $\hil$ equipped with the compact-open topology. A \emph{projective bundle over $X$} is a locally trivial bundle of projective spaces, i.e., a fibre bundle $P\to X$ with fibre $\PP(\hil)$ and structure group $PU(\hil)$. An application of the Banach-Steinhaus theorem shows that we may identify projective bundles with principal $PU(\hil)$-bundles (and the pointwise convergence topology on~$PU(\hil)$).

If $G$ is a topological group, let $G_X$ denote the sheaf of germs of continuous functions $G\to X$, i.e., the sheaf associated to the constant presheaf given by $U\mapsto F(U)=G$. Given a projective bundle $P\to X$ and a sufficiently fine good open cover $\{U_i\}_{i\in I}$ of $X$, the transition functions between trivializations $P|_{U_i}$ can be lifted to bundle isomorphisms $g_{ij}$ on double intersections $U_{ij}=U_i\cap U_j$ which are projectively coherent, i.e., over each of the triple intersections $U_{ijk}=U_i\cap U_j\cap U_k$ the composition $g_{ki}\,g_{jk}\,g_{ij}$ is given as multiplication by a $U(1)$-valued function $f_{ijk}:U_{ijk}\to U(1)$. The collection $\{(U_{ij},f_{ijk})\}$ defines a $U(1)$-valued two-cocycle called a $B$-field on $X$, which represents a class $B_P$ in the sheaf cohomology group $\RH^2(X,U(1)_X)$. On the other hand, the sheaf cohomology $\RH^1(X,PU(\hil)_X)$ consists of isomorphism classes of principal $PU(\hil)$-bundles, and we can consider the isomorphism class $[P]\in\RH^1(X,PU(\hil)_X)$. There is an isomorphism $\RH^1(X,PU(\hil)_X)\xrightarrow{\approx}\RH^2(X,U(1)_X)$ provided by the boundary map $[P]\mapsto B_P$. There is also an isomorphism 
$$
\RH^2\big(X\,,\,U(1)_X\big)~\xrightarrow{\approx}~
\RH^3(X,\bbz_X)\cong\RH^3(X,\bbz) \ .
$$

The image $\delta(P)\in\RH^3(X,\bbz)$ of $B_P$ is called the Dixmier-Douady invariant of $P$. When $\delta(P)=[H]$ is represented in $\RH^3(X,\bbr)$ by a closed three-form $H$ on $X$, called the $H$-flux of the given $B$-field $B_P$, we will write $P=P_H$. One has $\delta(P)=0$ if and only if the projective bundle $P$ comes from a vector bundle $E\to X$, i.e., $P=\PP(E)$. By Serre's theorem every torsion element of $\RH^3(X,\bbz)$ arises from a finite-dimensional bundle $P$. Explicitly, consider the commutative diagram of exact sequences of groups given by
\beq
\xymatrix{  0 \ar[r] & \mathbb{Z}_n \ar[r]  \ar[d] & SU(n) \ar[r]  \ar[d] & PU(n) \ar[r]  \ar[d]&  0\\
    0  \ar[r] & U(1)  \ar[r] & U(n)  \ar[r] & PU(n) \ar[r] & 0 \ , }
\label{SUnseq}\eeq
where we identify the cyclic group $\mathbb{Z}_n$ with the group of $n$-th roots of
unity. Let $P$ be a projective bundle with structure group $PU(n)$,
i.e., with fibres $\mathbb{P}(\mathbb{C}^n)$. Then the commutative
diagram of long exact sequences of sheaf cohomology groups associated to the commutative
diagram (\ref{SUnseq}) of groups implies that the element $B_P\in \RH^2(X,U(1)_X)$ comes from $\RH^2(X,(\mathbb{Z}_n)_X)$, and therefore its order divides $n$.

One also has $\delta(P_1\otimes P_2)=\delta(P_1)+\delta(P_2)$ and $\delta(P^\vee\,)=-\delta(P)$. This follows from the commutative diagram
$$
\xymatrix{  0 \ar[r] & U(1) \times U(1) \ar[r]  \ar[d] & U(\hil_1 ,\hil_2) \ar[r]  \ar[d] & PU(\hil_1 ,\hil_2) \ar[r]  \ar[d]&  0\\
    0  \ar[r] & U(1)  \ar[r] & U(\hil_1 \otimes \hil_2)  \ar[r] & PU(\hil_1 \otimes \hil_2) \ar[r] & 0 \ , }$$
and the fact that $P^\vee\otimes P=\PP(E)$ where $E$ is the vector bundle of Hilbert-Schmidt endomorphisms of $P$. Putting everything together, it follows that the cohomology group $\RH^3(X,\bbz)$ is isomorphic to the group of stable equivalence classes of principal $PU(\hil)$-bundles $P\to X$ with the operation of tensor product.

We are now ready to define the twisted K-theory of the manifold $X$ equipped with a projective bundle $P\to X$, such that $P_x=\PP(\hil)$ for all $x\in X$. We will first give a definition in terms of Fredholm operators, and then provide some equivalent, but more geometric definitions. Let $\hil$ be a $\bbz_2$-graded Hilbert space. We define $\Fred^0(\hil)$ to be the space of self-adjoint degree~1 Fredholm operators $T$ on $\hil$ such that $T^2-1\in\cK(\hil)$, together with the subspace topology induced by the embedding $\Fred^0(\hil)\hookrightarrow \balg(\hil)\times\cK(\hil)$ given by $T\mapsto (T,T^2-1)$ where the algebra of bounded linear operators $\balg(\hil)$ is given the compact-open topology and the Banach algebra of compact operators $\cK=\cK(\hil)$ is given the norm topology.

Let $P=P_H\to X$ be a projective Hilbert bundle. Then we can construct an associated bundle $\Fred^0(P)$ whose fibres are $\Fred^0(\hil)$. We define the \emph{twisted K-theory group of the pair $(X,P)$} to be the group of homotopy classes of maps
\beq
\K^0(X,H)=\big[X\,,\,\Fred^0(P_H)\big] \ .
\label{TwistedKproj}\eeq
The group $\K^0(X,H)$ depends functorially on the pair $(X,P_H)$, and an isomorphism of projective bundles $\rho:P\to P'$ induces a group isomorphism $\rho_*:\K^0(X,H)\to\K^0(X,H'\,)$. Addition in $\K^0(X,H)$ is defined by fibrewise direct sum, so that the sum of two elements lies in $\K^0(X,H_2)$ with $[H_2]=\delta(P\otimes\PP(\bbc^2))=\delta(P)=[H]$. Under the isomorphism $\hil\otimes\bbc^2\cong\hil$, there is a projective bundle isomorphism $P\to P\otimes\PP(\bbc^2)$ for any projective bundle $P$ and so $\K^0(X,H_2)$ is canonically isomorphic to $\K^0(X,H)$. When $[H]$ is a non-torsion element of $\RH^3(X,\bbz)$, so that $P=P_H$ is an infinite-dimensional bundle of projective spaces, then the index map $\K^0(X,H)\to\bbz$ is zero, i.e., any section of $\Fred^0(P)$ takes values in the index zero component of $\Fred^0(\hil)$.

Let us now describe some other models for twisted K-theory which will be useful in our physical applications later on. A definition in algebraic K-theory may given as follows. A bundle of projective spaces $P$ yields a bundle $\End(P)$ of algebras. However, if $\hil$ is an infinite-dimensional Hilbert space, then one has natural isomorphisms $\hil\cong\hil\oplus\hil$ and
$$
\End(\hil)~\cong~\Hom(\hil\oplus\hil,\hil)~\cong~\End(\hil)\oplus
\End(\hil)
$$
as left $\End(\hil)$-modules, and so the algebraic K-theory of the algebra $\End(\hil)$ is trivial. Instead, we will work with the Banach algebra $\cK(\hil)$ of compact operators on $\hil$ with the norm topology. Given that the unitary group $U(\hil)$ with the compact-open topology acts continuously on $\cK(\hil)$ by conjugation, to a given projective bundle $P_H$ we can associate a bundle of compact operators $\bun_H\to X$ given by
$$
\bun_H=P_H\times_{PU}\cK
$$
with $\delta(\bun_H)=[H]$. The Banach algebra $\alg_H:=C_0(X,\bun_H)$ of continuous sections of $\bun_H$ vanishing at infinity is the continuous trace $C^*$-algebra $CT(X,H)$~\cite{Rosenberg1}. Then the twisted K-theory group $\K^\bullet(X,H)$ of $X$ is canonically isomorphic to the algebraic K-theory group $\K_\bullet(\alg_H)$.

We will also need a smooth version of this definition. Let $\alg_H^\infty$ be the smooth subalgebra of $\alg_H$ given by the algebra $CT^\infty(X,H)=C^\infty(X,\calL_{P_H}^1)$, where $\calL_{P_H}^1=P_H\times_{PU}\calL^1$. Then the inclusion $CT^\infty(X,H)\hookrightarrow CT(X,H)$ induces an isomorphism $\K_\bullet\big(CT^\infty(X,H)\big)\xrightarrow{\approx}\K_\bullet\big(CT(X,H)\big)$ of algebraic K-theory groups. Upon choosing a bundle gerbe connection~\cite{Murray1,BCMMS1}, one has an isomorphism $\K_\bullet\big(CT^\infty(X,H)\big)\cong\K^\bullet(X,H)$ with the twisted K-theory (\ref{TwistedKproj}) defined in terms of projective Hilbert bundles $P=P_H$ over $X$.

Finally, we propose a general definition based on K-theory with coefficients in a sheaf of rings. It parallels the bundle gerbe approach to twisted K-theory~\cite{BCMMS1}. Let $\balg$ be a Banach algebra over $\bbc$. Let $\ecat(\balg,X)$ be the category of continuous $\balg$-bundles over $X$, and let $C(X,\balg)$ be the sheaf of continuous maps $X\to \balg$. The ring structure in $\balg$ equips $C(X,\balg)$ with the structure of a sheaf of rings over $X$. We can therefore consider left (or right) $C(X,\balg)$-modules, and in particular the category $\lfcat\big(C(X,\balg)\big)$ of locally free $C(X,\balg)$-modules. Using the section functor in the usual way, for $X$ compact there is an equivalence of additive categories
\beq
\ecat(\balg,X)\cong\lfcat\big(C(X,\balg)\big) \ .
\label{elfcatequiv}\eeq

Since these are both additive categories, we can apply the Grothendieck functor to each of them and obtain the abelian groups $\K(\lfcat(C(X,\balg)))$ and $\K(\ecat(\balg,X))$. The equivalence of categories (\ref{elfcatequiv}) ensures that there is a natural isomorphism of groups
\beq
\K\big(\lfcat\big(C(X,\balg)\big)\big)\cong
\K\big(\ecat(\balg,X)\big) \ .
\label{elfGrothiso}\eeq
This motivates the following general definition. If $\alg$ is a sheaf of rings over $X$, then we define the \emph{K-theory of $X$ with coefficients in $\alg$} to be the abelian group
$$
\K(X,\alg):=\K\big(\lfcat(\alg)\big) \ .
$$
For example, consider the case $\balg=\bbc$. Then $C(X,\bbc)$ is just the sheaf of continuous functions $X\to\bbc$, while $\ecat(\bbc,X)$ is the category of complex vector bundles over $X$. Using the isomorphism of K-theory groups (\ref{elfGrothiso}) we then have
$$
\K\big(X\,,\,C(X,\bbc)\big)~:=~
\K\big(\lfcat\big(C(X,\bbc)\big)\big)~\cong~
\K\big(\ecat(\bbc,X)\big)\= \K^0(X) \ .
$$

The definition of twisted K-theory uses another special instance of this general construction. For this, we define an \textit{Azumaya algebra over $X$ of rank $m$} to be a locally trivial algebra bundle over $X$ with fibre isomorphic
to the algebra of $m \times m$ complex matrices over $\mathbb{C}$,
$\mat_m(\mathbb{C})$. An example is the algebra $\End(E)$ of
endomorphisms of a complex vector bundle $E \rightarrow X$.
We can define an equivalence relation on the set $A(X)$ of Azumaya
algebras over $X$ in the following way. Two Azumaya algebras $A$,
$A{'}$ are called equivalent if there are vector bundles $E$,
$E{'}$ over $X$ such that the algebras $A \otimes \End(E)$, $A{'}
\otimes \End(E{'}\,)$ are isomorphic. Then every Azumaya algebra of
the form $\End(E)$ is equivalent to the algebra of functions $C(X)$ on $X$. The set of all equivalence
classes is a group under the tensor product of algebras, called the
\textit{Brauer group of $X$} and denoted $\Br(X)$. By Serre's theorem there is an isomorphism
$$
\delta \,:\, \Br(X) ~\xrightarrow{\approx}~ \tor\big(\RH^3(X,\mathbb{Z})\big) \ ,
$$
where $\tor(H^3(X,\mathbb{Z}))$ is the torsion subgroup of
$\RH^3(M,\mathbb{Z})$. For an explicit cocycle description of the Dixmier-Douady invariant $\delta (A)$ for an Azumaya algebra $A$, see ref.~\cite{Kapustin1}.

If $A$ is an Azumaya algebra bundle, then the space of continuous sections $C(X,A)$ of $X$ is a ring and we can
consider the algebraic K-theory group $\K(A):=\K_0(C(X,A))$ of equivalence
classes of projective $C(X,A)$-modules, which depends only on the
equivalence class of $A$ in the Brauer group~\cite{DoKa1}. Under the equivalence (\ref{elfcatequiv}), we can represent the Brauer group $\Br(X)$ as the set of isomorphism classes of sheaves of Azumaya algebras. Let $\alg$ be a sheaf of Azumaya algebras, and $\lfcat(\alg)$ the category of locally free $\alg$-modules. Then as above there is an isomorphism
$$
\K\big(X\,,\,C(X,\alg)\big)\cong\K\big({\rm Proj}\big(C(X,\alg)\big)\big) \ ,
$$
where ${\rm Proj}(C(X,\alg))$ is the category of finitely-generated projective $C(X,\alg)$-modules. The group on the right-hand side is the group $\K(A)$. For given $[H]\in\tor(\RH^3(X,\bbz))$ and $A\in\Br(X)$ such that $\delta(A)=[H]$, this group can be identified as the twisted K-theory group $\K^0(X,H)$ of $X$ with twisting $A$. This definition is equivalent to the description in terms of bundle gerbe modules, and from this construction it follows that $\K^0(X,H)$ is a subgroup of the ordinary K-theory of $X$. If $\delta(A)=0$, then $A$ is equivalent to $C(X)$ and we have $\K(A):=\K_0(C(X))=\K^0(X)$. The projective
$C(X,A)$-modules over a rank $m$ Azumaya algebra $A$ are vector
bundles $E \rightarrow X$ with fibre $\mathbb{C}^{n\,m} \cong 
(\mathbb{C}^m)^{\oplus n}$, which is naturally an $\mat_m(\mathbb{C})$-module. This
is a projective module and all projective $C(X,A)$-modules arise in
this way~\cite{Kapustin1}.

We will now describe the connection to twisted cohomology, following refs.~\cite{BCMMS1,MaSt1}. Upon choosing a bundle gerbe connection, one has an isomorphism of $\bbz_2$-graded cohomology groups
$$
\HP_\bullet\big(CT^\infty(X,H)\big)~\cong~\RH^\bullet(X,H)\=\RH^\bullet\big(\Omega^\bullet(X)\,,\,\dd-H\wedge\big)
$$
where the right-hand side is the $H$-twisted cohomology of $X$. The Chern-Weil representative, in terms of differential forms on $X$, of the canonical Connes-Chern character
$$
\ch\,:\,\K_\bullet \big( {CT}^\infty(X, H)\big) ~
\longrightarrow~ \HP_\bullet \big( {CT}^\infty(X, H)\big)
$$
then leads to the twisted Chern character
$$
{\ch}_H \,:\, \K^\bullet(X, H) ~\longrightarrow~ \RH^\bullet (X, H) \ .
$$

\subsection{Isometric pairing formulas}
\label{sec:7.2}

The Clifford algebra bundle $\Cl(T^*X)$ is an Azumaya algebra over $X$ with Dixmier-Douady invariant $\delta\big(\Cl(T^*X)\big)=w_3(X)$, the third Stiefel-Whitney class of the tangent bundle of $X$~\cite{Plymen1}. Consider the algebra
$$
\balg_H~:=~CT\big(X\,,\,w_3(X)-H\big)~\cong~ 
C_0\big(X\,,\,\bun_{-H}\otimes\Cl(T^*X)\big) \ .
$$
Then $(\alg_H,\balg_H)$ is a PD pair with fundamental class $\Delta=\Dirac\otimes\Dirac$~\cite{BMRS2,Tu1}. The restriction of the algebra $\alg_H\otimes\balg_H$ to the diagonal of $X\times X$ is isomorphic to the algebra $CT\big(X\,,\,w_3(X)\big)\otimes\cK$, which is Morita equivalent to the algebra of continuous sections $C_0\big(X\,,\,\Cl(T^*X)\big)$.

Under the isomorphism $\K^0\big(X\,,\,w_3(X)\big)\cong\K_0\big(C_0(X,\Cl(T^*X))\big)$, the tensor product of projective bundles defines a bilinear pairing on twisted K-theory groups given by
$$
\K^\bullet(X,H)\otimes\K^\bullet\big(X\,,\,w_3(X)-H\big)~
\longrightarrow~\K^0\big(X\,,\,w_3(X)\big)~\xrightarrow{\Ind}~\bbz \ .
$$
On the other hand, since the torsion class $w_3(X)$ is trivial in de~Rham cohomology, there is an isomorphism $\RH^\bullet\big(X\,,\,w_3(X)\big)\cong\RH^\bullet(X,\bbR)$ and hence the cup product defines a bilinear pairing on twisted cohomology groups via the mapping
$$
\RH^\bullet(X,H)\otimes\RH^\bullet\big(X\,,\,w_3(X)-H\big)~
\longrightarrow~\RH^{\rm even}(X,\bbR) \ .
$$
The fundamental cyclic cohomology class $\Xi$ of the PD pair $(\alg_H,\balg_H)$ may thus be identified with the orientation cycle $[X]$.

In this case the Todd class $\Todd(\alg_H)$ may be identified with the Atiyah-Hirzebruch genus $\widehat{A}(X)$ of the tangent bundle $TX$, and the modified Chern character (\ref{modch}) is $\ch_H\wedge\sqrt{\widehat{A}(X)}$. Note that when $X$ is a spin$^c$ manifold, then $w_3(X)=0$ and the algebra $C_0\big(X\,,\,\Cl(T^*X)\big)$ is Morita equivalent to $C(X)$~\cite{Plymen1}. In this instance $\balg_H=CT(X,-H)=\alg_H^\op$ is the opposite algebra of $\alg_H$, and the restriction of $\alg_H\otimes\balg_H$ to the diagonal of $X\times X$ is stably isomorphic to the algebra of functions $C(X)$.

\subsection{Twisted K-cycles and Ramond-Ramond charges}
\label{sec:7.3}

If spacetime $X$ is a spin manifold, then any D-brane $(W,E,f)$ in $X$ determines canonical element~\cite{CW1}
$$
f!\in\KK_d\big(CT(W,f^*[H]+w_3(\nu_W))\,,\,CT(X,H)\big) \ .
$$
Since $w_3(\nu_W)=w_3(W)$ in this case~\cite{Witten1,FW1}, we may identify the D-brane algebra $\dalg=CT\big(W\,,\,f^*[H]+w_3(W)\big)$ and the corresponding Chan-Paton bundle is an element $E\in\K^0\big(W\,,\,f^*[H]+w_3(W)\big)$. There are two particularly interesting special classes of such twisted D-branes.

The first class is determined by the usual requirement that the worldvolume $W$ be a spin$^c$ manifold, as in the ordinary Baum-Douglas construction. This instance was first considered in~ref.~\cite{MaSi1}. Then $w_3(W)=0$, the algebra $\dalg$ is the restriction of $\alg_H$ to $W$, and $E\in\K^0(W,f^*[H])$. The geometric K-homology equivalence relations are then completely analogous to those of the untwisted case in Section~\ref{sec:2.1}~\cite{MaSi1}. When the $H$-flux defines a non-torsion element in $\RH^3(X,\bbz)$, the Chan-Paton bundle $E$ is a projective bundle of infinite rank, corresponding to an infinite number of wrapped branes on $W$. When $H$ defines an $n$-torsion element, then the $B$-field $B_{P_H}$ incorporates the contribution from the $\bbz_n$-valued 't~Hooft flux necessary for anomaly cancellation on the finite system of $n$ spacetime-filling branes and antibranes in the $H$-flux background~\cite{Kapustin1}.

The second class is in some sense opposite to the first one, and it is more physical in that it is tied to the Freed-Witten anomaly cancellation condition~\cite{FW1}
\beq
f^*[H]+w_3(W)\=0 \ .
\label{FWcanc}\eeq
In this case $E\in\K^0(W)$ and the D-brane algebra $\dalg$ is (stably) commutative. The mathematical meaning of this limit is that it makes the worldvolume $W$ into a ``twisted spin$^c$'' manifold, which may be defined precisely as follows.
By Kuiper's theorem, the unitary group $U(\hil)$ of an infinite-dimensional
Hilbert space $\hil$ is contractible (both in the norm and compact-open
topologies). Thus the projective unitary group $PU(\hil)$ has the homotopy
type of an Eilenberg-Maclane space $K(\mathbb{Z},2)$, and its
classifying space $BPU(\hil)$ is an Eilenberg-Maclane space $K(\mathbb{Z},3)$. It follows that any
element of $\RH^3(X,\mathbb{Z})$ corresponds to a map $F : X
\rightarrow BPU(\hil)$, and hence to the projective bundle which is
the pullback by $F$ of the universal bundle over
$BPU(\hil)$.

It follows that $K(\mathbb{Z},3)$ is a classifying space for
the third cohomology, 
$$
\RH^3(X,\mathbb{Z}) \cong \big[X\,,\,K(\mathbb{Z},3)\big] \ ,
$$
and so we can represent an $H$-flux by a continuous map $H: X
\rightarrow K(\mathbb{Z},3)$. Taking a universal
$K(\mathbb{Z},2)$-bundle over $K(\mathbb{Z},3)$ and pulling it back
through $H$ to $X$, we get a $K(\mathbb{Z},2)$-bundle $P_H$ over
$X$. Consider the $K(\mathbb{Z},2)$-bundle
\begin{eqnarray*}
& BU(1) & ~\longrightarrow ~BSpin^c~ \longrightarrow~ BSO \\
& \parallel & \\ & K(\mathbb{Z},2) &
\end{eqnarray*}
with classifying map $\beta\circ w_2 : BSO \rightarrow
BBU(1)=K(\mathbb{Z},3)$, the Bockstein homomorphism of the second Stiefel-Whitney class. The action of $K(\mathbb{Z},2)$ on $BSpin^c$ induces a
principal $BSpin^c$-bundle $Q=P_H\times_{K(\mathbb{Z},2)}BSpin^c$,
i.e., a sequence of bundles $Q_n=P_H\times_{K(\mathbb{Z},2)}BSpin^c(n)$,
with corresponding universal bundles
$UQ_n=(P_H\times_{K(\mathbb{Z},2)}ESpin^c(n))
\times_{Spin^c(n)}\mathbb{R}^n$. The homotopy groups of the
associated Thom spectrum
$$
{\rm Thom}(UQ)=P_+\,\bigwedge_{K(\mathbb{Z},2)_+}\,MSpin^c
$$
are the $H$-twisted spin$^c$ bordism groups of $X$.
Using this one can deduce that a compact manifold $W$ is $H$-twisted $\K$-oriented if
it is an oriented manifold with a continuous map $f : W
\rightarrow X$ such that the Freed-Witten condition (\ref{FWcanc}) holds. We say that a pair
$(W,f)$, with $W$ a compact oriented manifold and $f: W
\rightarrow X$ a continuous map, is $H$-twisted spin$^c$ if
it satisfies this cancellation. 

A choice of $H$-twisted
spin$^c$ structure is a choice of a two-cochain $c$ such that, at
the cochain level, $\delta(c)=\beta\circ w_2(W)-f^*[H]$. This follows from the following geometric fact. Let $\alpha : P \rightarrow P$ be an automorphism of a projective bundle $P\to X$ with infinite-dimensional separable fibres. It induces a line bundle $L_{\alpha} \rightarrow X$. For $x \in X$, the
non-zero elements of $(L_{\alpha })_x$ are the linear isomorphisms
$E_x \rightarrow E_x$ which induce $\alpha |_{P_x}$, where
$P_x=\mathbb{P}(E_x)$. Then the assignment $\Aut(P)
\rightarrow \RH^2(X,\mathbb{Z}), \, \alpha \mapsto [L_{\alpha }]$
identifies the group of connected components $\pi_0(\Aut(P))$ with the
group $\RH^2(X,\mathbb{Z})$ of isomorphism classes of line bundles
over $X$. This follows from the identification of the automorphism group $\Aut(P)$ of the bundle $P$ with the space of sections of the endormorphism bundle $\End(P)$, i.e., the space of maps $X \rightarrow PU(\hil)$, which is an Eilenberg-Maclane space $K(\mathbb{Z},2)$. For
a more extensive treatment of these issues, see refs.~\cite{Dou1,Wang1}. 

This leads us to the following notion. Let $(W,f)$ be a manifold (not necessarily $H$-twisted
spin$^c$). A vector bundle $V \rightarrow W$ is said to be an
\textit{$H$-twisted spin$^c$ vector bundle} if $f^*[H]=
w_3(V)$. The choice of a specific $H$-twisted spin$^c$ structure on $V$
is made as above by choosing an appropriate two-cochain. The notion of an $H$-twisted spin$^c$ manifold is
just the special case $V=TW$ of this latter one. The analogs of the Baum-Douglas gauge equivalence relations for geometric twisted K-homology may be straightforwardly written down in the obvious way using projective Hilbert bundles instead of vector bundles. In the construction of the unit sphere bundle (\ref{unitspherebun}), we assume that $w_3(\,\widehat{W}\,)=\pi^*(f^*[H])$. Then $(\,\widehat{W},f\circ\pi)$ is an $H$-twisted spin$^c$ manifold. The rest of the construction proceeds by using the untwisted Thom class $H(F)\in\K^i(\,\widehat{W}\,)$. See ref.~\cite{Wang1} for the relation to a description involving bundle gerbe modules.

There are more general twistings one may consider which are still
physically meaningful. Suppose that $[H]\in \mathbb{Z}_n \subset
\RH^3(X,\mathbb{Z})$ and fix an element $y \in \RH^3(X,\mathbb{Z}_n)$. Then we may
consider bordism of manifolds $(W,f)$, where the worldvolume $W$ is a compact
oriented manifold and $f:W\rightarrow X$ is a continuous map
satisfying
\beq
f^*[H]= w_3(W)+f^*\big(\beta(y)\big) \ , 
\label{genFWcond}\eeq
with $\beta$ the Bockstein homomorphism. The condition (\ref{genFWcond}) is the most general form of the Freed-Witten anomaly
cancellation condition for a system of $n$ spacetime-filling brane-antibrane pairs~\cite{Kapustin1}.
With this more general kind of twisting, one can also consider bordism of manifolds $(W,f)$, where $W$ is a compact
spin$^c$ manifold as before and $f:W\rightarrow X$ is a continuous map
satisfying $f^*[H]=f^*(\beta(y))$. The equivalences between the various forms of the geometric twisted K-homology group $\K_\bullet(X,H)$ follows from the equivalences among the corresponding twisted K-theories. In any of these cases, one arrives at the twisted D-brane charge vector
$$
{Q_H(W,E,f)\=\ch_H\big(f_!(E)\big)\wedge\sqrt{\widehat{A}(X)}}\in\RH^\bullet(X,H) \ .
$$
Only when $[H]$ is a torsion class does the Ramond-Ramond charge correspond to an element of the ordinary (untwisted) cohomology of the spacetime manifold $X$.

\section{Correspondences and T-duality}
\label{sec:8}

In this final section we shall apply our formalism to a new description of topological open string T-duality~\cite{BMRS1,BMRS2}. The description is based on the formulation of KK-theory in terms of correspondences~\cite{CSk1,BW1,Cuntz1}. Amongst other things, this leads to an explicit construction of the various structures inherent in Kasparov's bivariant K-theory, and moreover admits a natural noncommutative generalization~\cite{BMRS2}.

\subsection{Correspondences}
\label{sec:8.1}

Let $X,Y$ be smooth manifolds, and set $\KK_d(X,Y):=\KK_d\big(C_0(X)\,,\,C_0(Y)\big)$. Elements of the group $\KK_d(X,Y)$ can be represented by \emph{correspondences}
$$
\xymatrix{ &(Z,E)\ar[ld]_f\ar[rd]^g& \\ X & & Y }
$$
where $Z$ is a smooth manifold, $E$ is a complex vector bundle over $Z$, the map $f:Z\to X$ is smooth and proper, $g:Z\to Y$ is a smooth K-oriented map, and $d=\dim(Z)-\dim(Y)$. This diagram defines a morphism 
$$
g_!\big(f^*(-)\otimes E\big)\in
\Hom\big(\K^\bullet(X)\,,\,\K^{\bullet+d}(Y)\big)
$$
implemented by the KK-theory class $[f]\otimes_{C_0(Z)}[[E]]\otimes_{C_0(Z)}(g!)$, where $[[E]]$ is the KK-theory class in $\KK_0(Z,Z)\cong\End\big(\K^\bullet(Z)\big)$ of the vector bundle $E$ defined by tensor product with the K-theory class $[E]$ of $E$ (this ignores the extension term in the universal coefficient theorem (\ref{UCTKK})). The collection of all correspondences forms an additive category under disjoint union. The group $\KK_d(X,Y)$ is then obtained as the quotient space of the set of correspondences by the equivalence relation generated by suitable notions of cobordism, direct sum and vector bundle modification, analogous to those of Section~\ref{sec:2.1}~\cite{BMRS2}.

The correspondence picture of KK-theory gives a somewhat more precise realization of the notion, introduced categorically in Section~\ref{sec:3.3}, of Kasparov bimodules as ``generalized'' morphisms of $C^*$-algebras. It provides a geometric presentation of the analytic index for families of elliptic operators on $X$ parametrized by $Y$. The limiting case $\KK_d(X,\pt)=\K_d(X)$ is the geometric K-homology of $X$ as described in Section~\ref{sec:2}, since in this case a correspondence is simply a Baum-Douglas K-cycle $(Z,E,f)$ over $X$. On the other hand, the group $\KK_d(\pt,Y)=\K^d(Y)$ is the K-theory of $Y$, obtained via an ABS-type construction of the charge of the D-brane $(Z,E,g)$ in $Y$ using the spin$^c$ structure on the bundle $TZ\oplus g^*(TY)$.

One of the great virtues of this formalism is that it gives an explicit description of the intersection product in KK-theory, which as mentioned in Section~\ref{sec:3.3} is notoriously difficult to define. In the notation above it is a map
$$
\otimes_M\,:\,\KK(X,M)\times\KK(M,Y)~\longrightarrow~
\KK(X,Y)
$$
which sends two correspondences
$$
\xymatrix{ &(Z_1,E_1)\ar[ld]_f\ar[rd]^{g_M}& &
(Z_2,E_2)\ar[ld]_{f_M}\ar[rd]^{g} & \\ X & & M & & Y }
$$
to the correspondence
$$
[Z,E]\=[Z_1,E_1]\otimes_M[Z_2,E_2]
$$
with $Z=Z_1\times_M Z_2$ and $E=E_1\boxtimes E_2$. To ensure that the fibred product $Z$ is a smooth manifold, one has to impose the transversality condition
$$
\dd f_M(T_{z_2} Z_2) + \dd g_M(T_{z_1} Z_1) \= T_{f_M(z_2)} M 
$$
for all $(z_1,z_2)\in Z_1\times Z_2$. Such choices can always be straightforwardly made using standard transversality theorems and homotopy invariance of the KK-functor, such that this restricted set of correspondences is in a sense ``dense'' in the space of all correspondences~\cite{CSk1}.

\subsection{T-duality and KK-equivalence}
\label{sec:8.2}

The correspondence picture is reminescent of the Fourier-Mukai transform, which is related to T-duality on spacetimes compactified on tori $X=M \times \bbT^n$
in the absence of a background $H$-flux. In this case the T-dual is topologically the same space $M \times \widehat\bbT^n$, and the mechanism implementing the T-duality is given by the smooth analog of the Fourier-Mukai transform~\cite{Hori1}. Let $\bbt^n$ be an $n$-torus, and let $\widehat{\bbt}{}^n\cong\Pic^0(\bbt^n)$ be the corresponding dual $n$-torus. Recall that the Poincar\'e line bundle $\Poin_0\rightarrow\bbt^n\times\widehat{\bbt}{}^n$ is the unique line bundle such that $\Poin_0\big|_{\bbt^n\times\{\,\widehat{t}~\}}\in\Pic^0(\bbt^n)$ is the flat line bundle corresponding to $\widehat{t}\in\widehat{\bbT}{}^n$ and whose restriction $\Poin_0\big|_{\{0\}\times\dtorus{}^n}$ is trivial.

This data defines a diagram
$$
\xymatrix{ &\big(M\times\bbt^n\times\dtorus{}^n\,,\,\Poin\big)
\ar[ld]_{p_1}\ar[rd]^{p_2}& \\ M\times\bbt^n & & M\times\dtorus{}^n }
$$
where $p_1,p_2$ are canonical projections and $\Poin$ is the pullback of the Poincar\'e line bundle to $M\times\bbt^n\times\dtorus{}^n$. The smooth analog of the Fourier-Mukai transform is the isomorphism of K-theory groups
$$
T_!\,:\,\K^\bullet\big(M\times\bbt^n\big)~\xrightarrow{\approx}~
\K^{\bullet+n}\big(M\times\dtorus{}^n\big)
$$
given by
$$
T_!(-)\=(p_2)_!\big(p_1^*(-)\otimes\Poin\big) \ .
$$
We conclude that \emph{topological open string T-duality} is a correspondence. In this case, the correspondence represents an invertible element of KK-theory, i.e., a KK-equivalence.

The Fourier-Mukai transform can be rephrased in a satisfactory manner,
entirely in terms of noncommutative geometry, as a crossed product algebra
${C}_0(M\times \bbT^n)\rtimes \bbR^n$, where
the action of the group $\bbR^n$ on ${C}_0(M\times \bbT^n)$ is just the given 
action of $\bbR^n$ on $\bbT^n$ by translations and the trivial action on $M$. By Rieffel's version of the Mackey imprimitivity theorem~\cite{Rieffel1}, 
one sees that the crossed product $C^*$-algebra ${C}_0(M\times \bbT^n)\rtimes \bbR^n$ is 
Morita equivalent to 
$$
C_0(M)\otimes C^*(\bbR^n)
\cong C_0\big(M\times\dtorus{}^n\big) \ .
$$
Thus the T-dual of the $C^*$-algebra ${C}_0(M\times \bbT^n)$ is 
obtained by taking the crossed product of the algebra with 
$\bbR^n$. The Connes-Thom isomorphism then defines a \emph{KK-equivalence}
$$
\alpha~\in~\KK_n\big(M\times\bbt^n\,,\,M\times\dtorus{}^n\big)
$$
which is just the families Dirac operator. Moreover, Takai duality gives a Morita equivalence
$$\big(C_0(M\times\bbt^n)\rtimes\bbR^n\big)\rtimes\bbR^n\sim
C_0(M\times\bbt^n) \ , 
$$
showing that the T-duality transformation is topologically of order~2.

The reason for making this reformulation in terms of noncommutative geometry is that it extends to the case when spacetime $X$ is a principal torus bundle $\pi:E\xrightarrow{\bbt^n}M$ of rank $n$ in the presence of a background $H$-flux. In this instance the T-dual is a crossed product algebra $CT(E,H)\rtimes\bbR^n$, which is generally a bundle of rank $n$ noncommutative tori fibred over $M$~\cite{MR1}. This requires that $H$ restrict to zero in the cohomology of the torus fibers and that the action of $\bbR^n$ on the continuous trace $C^*$-algebra ${CT}(X, H)$ is a lift of the given action of $\bbR^n$ on $X$. That such a lift exists is a non-trivial result proven in ref.~\cite{MR1}. This crossed product algebra is a noncommutative $C^*$-algebra, but it need not be a continuous trace algebra. In ref.~\cite{GSN1} it was shown, by checking the open string metric, that in some cases these algebras are globally defined, open string versions of T-folds. The correspondence picture in this context appears to nicely describe the doubled torus formalism for T-folds, as we will see below. When $\pi_*[H]=0$, the T-dual algebra is isomorphic to a continuous trace $C^*$-algebra $CT\big(\,\widehat{E}\,,\,\widehat{H}\,)$ and represents a geometrically dual spacetime in the usual sense.

\subsection{Noncommutative correspondences}
\label{sec:8.3}

The discussion at the end of Section~\ref{sec:8.2} above motivates the following noncommutative generalization of the correspondence picture of Section~\ref{sec:8.1} above~\cite{BMRS2}. Let $\alg,\balg$ be separable $C^*$-algebras. We will represent elements of $\KK(\alg,\balg)$ by \emph{noncommutative correspondences}
$$
\xymatrix { & (\calg , \xi) 
& \\
\alg\ar[ur]^{f}  &   &    \balg\ar[ul]_{g} }
$$
where $\calg$ is a separable $C^*$-algebra and $\xi \in \KK(\calg, \calg)$, whereas $f:\alg\to\calg$ and $g:\balg\to\calg$ are homomorphisms with $g$ $\K$-oriented. The intersection product gives an element $[f]\otimes_\calg \xi \otimes_\calg (g!) \in\KK(\alg, \balg)$, with associated K-theory morphism $g^!\big(f_*(-) \otimes_\calg
\xi\big)\in\Hom\big(\K_\bullet( \alg)\,,\, 
\K_{\bullet}(\balg)\big)$.
Every class in $\KK_d(\alg,\balg)$ comes from a noncommutative correspondence, in fact from one with trivial $\xi=1_\calg$.
The representation of the intersection product in this instance uses amalgamated products of $C^*$-algebras~\cite{BMRS2}.

Let us consider the class of examples mentioned earlier, focusing for simplicity on the simplest case where spacetime $X$ is a principal circle bundle $\pi:E\xrightarrow{\bbt}M$ in a background $H$-flux. The T-dual is another principal circle bundle $\widehat{\pi}:\widehat{E}\xrightarrow{\widehat{\bbt}}M$ with characteristic class $c_1(\widehat{E})=\pi_*[H]$. The Gysin sequence for $E$ defines the T-dual $H$-flux $[\,\widehat{H}\,]\in \RH^3\big(\,\widehat{E}\,,\,\bbz\big)$ with $c_1(E) = \widehat{\pi}_*[\, \widehat{H}\,]$ and $[H] = [\,\widehat{H}\,]$ in $\RH^3\big(E\times_M \widehat{E}\,,\,\bbz\big)$. This data defines a noncommutative correspondence
$$
\xymatrix{ & \big(CT(E\times_M \widehat{E},H) \,,\, \xi\big) 
& \\
CT(E, H)\ar[ur]^{f}  &   &    CT\big(\widehat{E}\,,\, 
\widehat{H}\,\big)\ar[ul]_{g} }
$$
where $\xi$ is an analogue of the Poincar\'e line bundle. It determines a KK-equivalence $\alpha\in\KK_1\big(CT(E, H)\,,\, CT(\,\widehat{E}, \widehat{H}\,)\big)$. See ref.~\cite{BMRS2} for further examples of noncommutative correspondences.

\subsection{Axiomatic T-duality and D-brane charge}
\label{sec:8.4}

Inspired by the above results, we now give an axiomatic definition of T-duality
in $\K$-theory that any definition of the \emph{T-dual} $T(\alg)$ of a $C^*$-algebra $\alg$ should satisfy. These axioms include the requirements that the Ramond-Ramond charges of $\alg$ should be in bijective correspondence with the Ramond-Ramond charges of ${T}(\alg)$, and that T-duality applied twice yields a $C^*$-algebra which is physically equivalent to the $C^*$-algebra that we started out with. For this, we postulate the existence of a suitable category of separable $C^*$-algebras, possibly with extra structure (for example the $\bbR^n$-actions used above). Its objects $\alg$ are called \emph{T-dualizable algebras} and satisfy the following requirements:
\begin{enumerate}
\item The map $\alg\mapsto T(\alg)$ from $\alg$ to the {T-dual} of $\alg$ is a covariant functor;
\item There is a functorial map $\alg\mapsto\alpha_\alg$, where the invertible element $\alpha_\alg$ defines a KK-equivalence in $\KK\big(\alg\,,\,T(\alg)\big)$; and
\item The algebras $\alg$, $T\big(T(\alg)\big)$ are Morita
  equivalent, with associated KK-equivalence given by the invertible element
  $\alpha_\alg\otimes_{T(\alg)}\alpha_{T(\alg)}$.
\end{enumerate}

Let us consider a class of examples generalizing those already presented in this section. Let $\alg$ be a $G$-$C^*$-algebra, where $G$ is a 
locally compact, abelian vector Lie group (basically $\bbR^n$).
Then the algebra ${T}(\alg) = \alg\rtimes G$ satisfies the axioms above~\cite{BMRS1}, thanks to the Connes-Thom isomorphism and Takai duality (here we 
tacitly identify $G$ with its Pontrjagin dual $\tilde G$). 
The assumption made above that the T-dual ${T}(\alg)$ 
is a $C^*$-algebra is very strong and it is not always 
satisfied, as seen in ref.~\cite{BHM1}. Yet even in that
case, the axioms above are satisfied, provided one also allows more general 
algebras belonging to a category studied there. There is also an analogous
axiomatic definition of T-duality in local cyclic cohomology~\cite{BMRS1}, relevant to the duality transformations of Ramond-Ramond fields.

A crucial point about the formulation in terms of bivariant K-theory is that it provides a \emph{refinement} of the usual notion of T-duality. For instance, for a suitable class of algebras the universal coefficient theorem (\ref{UCTKK}) expresses the KK-theory group $\KK_\bullet(\alg,\balg)$ as an extension of the group $\Hom_\bbz\big(\K_\bullet(\alg)\,,\,\K_\bullet(\balg)\big)$ by $\Ext_\bbz\big(\K_{\bullet+1}(\alg)\,,\,\K_\bullet(\balg)\big)$. The extension group can lead to important torsion effects not present in the usual formulations of T-duality.

We close by studying the invariance of the noncommutative D-brane charge vector (\ref{NCRRcharge})  under T-duality. As is well known~\cite{Myers1}, the T-duality invariance of Ramond-Ramond couplings on D-branes is a subtle issue which requires further conditions to be imposed on the structures involved. The present formalism yields a systematic and general way to establish these criteria.

If the D-brane algebra $\dalg$ is a PD algebra, then by the Grothendieck-Riemann-Roch formula (\ref{NCGRR}) one has
$$
{Q}(\dalg,\xi,f) \= \ch(\xi)  \otimes_\dalg
\,\Todd(\dalg)\otimes_\dalg (f*)
\otimes_\alg\sqrt{{\Todd}(\alg)}\,^{-1} \ .
$$
Suppose that there is a local cyclic cohomology class $\Lambda\in\HL(\dalg,\dalg)$
 such that
$$
{(f*)\otimes_\alg\sqrt{{\Todd}(\alg)}\,^{-1}\=\Lambda\otimes_\dalg(f*)}  \ .
$$
Then there is a noncommutative version of the Wess-Zumino class (\ref{WZclass}) in $\HL_\bullet(\dalg)$ given by
$$
{D}_{\rm WZ}(\dalg,\xi,f)\=\ch(\xi)  \otimes_\dalg
\,\Todd(\dalg)\otimes_\dalg\Lambda \ .
$$

Consider a pair of D-branes $(\dalg,\xi,f)$ and $(\dalg',\xi',f'\,)$ which are $\KK$-equivalent, with the equivalence determined by an invertible element $\alpha$ in $\KK(\dalg,\dalg'\,)$ and $\xi'=\xi\otimes_\dalg\alpha$. If
$$
{\Lambda'\=\ch(\alpha)^{-1}\otimes_\dalg\Lambda\otimes_\dalg
\,\ch(\alpha)}
$$
then by eq.~(\ref{ToddKKrel}) one has ${D}_{\rm WZ}(\dalg',\xi',f'\,)={D}_{\rm WZ}(\dalg,\xi,f)\otimes_\dalg \,\ch(\alpha)$. It follows that
$$
{D}_{\rm WZ}(\dalg'',\xi'',f''\,)\={D}_{\rm WZ}(\dalg,\xi,f)\otimes_\dalg
\,\ch(\alpha\otimes_{\dalg'}\alpha'\,)
$$
in $\HL_\bullet(\dalg''\,)\cong\HL_\bullet(\dalg)$. This formula expresses the desired T-duality covariance under the conditions spelled out above.

\subsection*{Acknowledgments}

The author would like to thank the organisors and participants of the
workshop for the very pleasant scientific and social atmosphere. He
would especially like to thank J.~Brodzki, V.~Mathai, R.~Reis,
J.~Rosenberg and A.~Valentino for the enjoyable collaborations and
extensive discussions over the last few years, upon which this article
is based. This work was supported in part by the Marie Curie Research
Training Network Grant {\sl ForcesUniverse} (contract
no.~MRTN-CT-2004-005104) from the European Community's Sixth Framework
Programme.

%
%

%
%



\end{document}